\documentclass[%
 reprint,
 amsmath,amssymb,
 aps,
 pra,
]{revtex4-2}

\usepackage{xcolor,soul,framed} 
\usepackage{graphicx}
\usepackage{dcolumn}
\usepackage{bm}
\usepackage{braket}
\usepackage{physics}
\usepackage{amsmath}
\usepackage{amsfonts}
\usepackage{amssymb}
\usepackage{float}
\usepackage{ulem}
\newcommand{\covar}{\bm{\gamma}}
\newcommand{\disp}{\vec{d}}
\newcommand{\dense}{\hat{\rho}}

\usepackage[caption=false]{subfig}
\usepackage{hyperref}

\newcommand{\DM}[1]{\textcolor{pink}{DM}} 

\begin{document}

\title{Analytical Modeling of Real-World Photonic Quantum Teleportation}

\author{Neil~Sinclair$^{1,2,3}$}
\author{Samantha~I.~Davis$^{1,2}$}
\author{Nikolai~Lauk$^{1,2}$}\altaffiliation{Present Address: Photonic Inc., BS, Canada}
\author{Chang~Li$^{1,2}$}\altaffiliation{Present Address: Joint Quantum Institute, University of Maryland, College Park, MD 20742, USA}
\author{Damian~R.~Musk$^{1,2}$}
\author{Kelsie~Taylor$^{1,2}$}
\altaffiliation{Present Address: Weinberg Institute for Theoretical Physics, University of Texas at Austin, Austin, TX 78712, USA}
\author{Raju~Valivarthi$^{1,2}$}
\author{Maria~Spiropulu$^{1,2}$}
 
\affiliation{$^1$Division of Physics, Mathematics and Astronomy, California Institute of Technology, 1200 E. California Blvd., Pasadena, CA 91125, USA}
\affiliation{$^2$Alliance for Quantum Technologies (AQT/INQNET), California Institute of Technology, 1200 E. California Blvd., Pasadena, CA 91125, USA}
\affiliation{$^3$John A. Paulson School of Engineering and Applied Science, Harvard University, 29 Oxford St., Cambridge, MA 02138, USA}

\date{\today}

\begin{abstract}
We develop analytical models for realistic photonic quantum teleportation experiments with time-bin qubits, utilizing phase space methods from quantum optics. These models yield analytical expressions for Hong-Ou-Mandel interference visibilities and qubit fidelities, accounting for imperfections such as loss, photon distinguishability, and unwanted multi-photon events in single and entangled photon sources. Our expressions agree with the Hong-Ou-Mandel interference visibilities and teleportation fidelities reported by Valivarthi et al. (PRX Quantum 1, 020317, 2020). We use these models to predict and analyze the outcomes of future teleportation experiments under varying degrees of imperfections.
\end{abstract}

\maketitle

\section{Introduction} Quantum teleportation \cite{bennett93} is a key process for distributing qubits in a quantum network \cite{kimble2008quantum, wehner:2018, simon:2017}. Optical photons are used for long-distance networking due to their high velocities, carrier frequencies, and ease of encoding, manipulation, transmission, and detection \cite{gisin2007quantum, valivarthi2016quantum, sun2016quantum, takesue2009entanglement}. However, achieving ideal single and entangled photons for quantum networks and other quantum information tasks remains challenging \cite{eisaman2011invited, o2009photonic}. Photonic quantum states are often approximated using states that are easier to generate, such as those produced with room-temperature off-the-shelf equipment \cite{wang2007quantum, weedbrook2012gaussian}. For example, Gaussian states like weak coherent and two-mode squeezed states can substitute single and biphoton states but introduce additional photons that cause errors in quantum networks \cite{christ2012limits}. Additional errors arise from the lack of control over all degrees of freedom of a photon and device imperfections. Photon loss further hinders network deployment over distances greater than tens of kilometers \cite{takeoka2014fundamental}. Therefore, it is crucial to account for sources of loss and errors to accurately model and predict the performance of quantum networking experiments, particularly quantum teleportation. Predicting experimental performance under various operating conditions, both in the lab and in real-world applications, is also important.

We use the phase space formulation of quantum optics to analytically model post-selected discrete-variable quantum teleportation experiments with realistic imperfections such as loss, photon distinguishability, and imperfect sources of single and entangled photons. Our model is based on a recent time-bin qubit quantum teleportation experiment \cite{valivarthi2020}, providing expressions for key figures-of-merit including Hong-Ou-Mandel (HOM) interference visibilities and teleportation fidelities as functions of these imperfections. This experiment's use of Gaussian states allows us to model transformations and imperfections with Gaussian operations~\cite{Takeoka:2015, Weedbrook:2012}, differing from the conventional "photon-by-photon" approach that becomes cumbersome with higher number states and experimental imperfections. We use the characteristic function formalism from the phase space representation~\cite{Takeoka:2015} to derive closed analytical expressions for these figures-of-merit. Our models show excellent agreement with the findings in Ref. \cite{valivarthi2020}. Finally, we project the performance of future quantum teleportation experiments with varying degrees of imperfections and discuss our findings, their impact on future work, and the limitations of our model.

\section{Characteristic function formalism}
\label{section:characteristicfunction}
\subsection{Phase space representation and characteristic function}
Any quantum state of an $n$-mode bosonic system with the corresponding creation and annihilation operators $\hat{a}^\dagger_k$ and $\hat{a}_k$ that obey commutator relations
\begin{align}
    \left[ \hat{a}_k, \hat{a}^\dagger_l\right]=\delta_{kl},
\end{align}
can be described using the quadrature operators \cite{weedbrook2012gaussian}
\begin{align}
    \hat{x}_l=\frac{1}{\sqrt{2}}\left(\hat{a}^\dagger_l+\hat{a}_l\right)\quad\textrm{and}\quad \hat{p}_l=\frac{i}{\sqrt{2}}\left(\hat{a}^\dagger_l-\hat{a}_l\right),
\end{align}
which satisfy the canonical commutator relations
\begin{align}
    \left[\hat{x}_l,\hat{p}_k\right]=i\delta_{kl}.
\end{align}
Using a Weyl operator $\hat{\mathcal{W}}(\vec{\xi})=\exp(-i\vec{\xi}\cdot\hat{\vec{R}})$ with $\hat{\vec{R}}=(\hat{x}_1,\hat{p}_1,\dots,\hat{x}_n,\hat{p}_n)$ being a vector of quadrature operators of individual modes, we can construct an invertible mapping between functions over phase space and operators over the Hilbert space of the bosonic systems. 
In particular we can assign a characteristic function to any quantum state that is described by a density matrix $\hat{\rho}$ as follows
\begin{align}
    \chi(\vec{\xi})=\mathrm{Tr}\left\{\hat{\rho}\hat{\mathcal{W}}(\vec{\xi})\right\}.\label{eq:char_func}
\end{align}

\subsection{Gaussian states and unitaries}
Photon number or Fock states, such as single photons or entangled two-photon states, are essential for many quantum networking schemes, including those relying on quantum teleportation. 
However, photon number states are highly non-classical and are difficult to produce on demand and with specific properties, such as  near-unity indistinguishability.  In many experiments, weak coherent states approximate single photons, and two-mode squeezed states, produced at the output of a bulk nonlinear optical crystal, substitute photon pairs \cite{eisaman2011invited}. Two-mode squeezed states are also used for heralded single photon sources \cite{hong1986experimental}. Besides their straightforward preparation, coherent and two-mode squeezed states have the convenient property of being Gaussian, meaning their characteristic function is given by a multivariate Gaussian function
\begin{align}
    \chi(\vec{\xi})=\exp(-\frac{1}{4}\vec{\xi}^T\gamma\vec{\xi}-i\vec{d}^T\vec{\xi}),
\end{align}
where $\gamma$ is the covariance matrix and $\vec{d}$ is the displacement vector. 
Examples of Gaussian states include single-mode vacuum $\ket{0}$ with $\gamma=\mathbb{I}$ and $\vec{d}=0$, coherent states $\ket{\alpha}$ with $\gamma=\mathbb{I}$ and $\vec{d}=\sqrt{2}(\operatorname{Re}(\alpha), \operatorname{Im}(\alpha))^T$, as well as thermal states $\hat{\rho}=\sum_n \frac{\mu^n}{(\mu+1)^{n+1}}\ket{n}\bra{n}$ with $\gamma=(1+2\mu)\mathbb{I}$ and $\vec{d}=0$.

\subsection{Gaussian representation of relevant operations in photonic quantum teleportation}
A unitary operation on the Hilbert space that transforms Gaussian states into Gaussian states is called Gaussian unitary. 
The action of a Gaussian unitary on a Gaussian state results in simply changing the covariance matrix and displacement vectors $\gamma \rightarrow \gamma' = S\gamma S^T$ and $d \rightarrow d' = Sd$, where $S$ the symplectic matrix corresponding to a Gaussian unitary \cite{weedbrook2012gaussian, Takeoka:2015}. 
Note also that $ \hat{\vec{R^\prime}}= S\hat{\vec{R}}$. 
Below we consider some examples of Gaussian operations that are relevant for modeling photonic quantum teleportation.

\subsubsection{Phase shifter operation}
The first example is the addition of a constant phase $\phi$ to the field state. 
Its action in operator space is described by the following transformation: $\hat a \rightarrow \mathrm{e}^{i\phi}\hat a$, and the corresponding symplectic matrix is given by
\begin{align}
    S=\begin{pmatrix}
\cos(\phi) & -\sin(\phi)\\
\sin(\phi) & \cos(\phi)
\end{pmatrix}.
\end{align}

\subsubsection{Beam splitter operation}
Another very frequently used element in experiments are two-port beam splitters. 
For a beam splitter with transmittance $t$ and reflectance $r$ the transformation between the input ($\hat{a}, \hat{b}$) modes and output ($\hat{c}, \hat{d}$) modes is described by
\begin{align}
    \begin{pmatrix}
        \hat{c}\\
        \hat{d}
    \end{pmatrix} = \begin{pmatrix}
        t & i r\\
        i r & t
    \end{pmatrix}\begin{pmatrix}
        \hat{a}\\
        \hat{b}
    \end{pmatrix},
\end{align}
and the corresponding symplectic transformation between the input and the output modes is
\begin{align}
     S=\begin{pmatrix}
t & 0 & 0 & -r\\
0 & t & r & 0\\
0 & -r & t & 0\\
r & 0 & 0 & t
\end{pmatrix}.
\end{align}

\subsubsection{Channel loss}
The overall channel transmission $\eta$, taking into account propagation loss, inefficient couplings and detectors, can be modeled by a mixing the mode of interest with a vacuum mode on a virtual beam splitter of transmittance $\sqrt{\eta}$ and tracing out the transmitted part of this vacuum mode. 
This results \cite{Takeoka:2015} in the following transformation of the covariance matrix and the displacement vector
\begin{align}
\tilde\gamma &= \eta \gamma + (1-\eta)\mathbb{I},\\
\vec{\tilde d} &= \sqrt{\eta} \vec{d}.
\end{align}

\subsubsection{Measurements}
Usually, the process of detection or photon counting is not a Gaussian process. 
However, in the case of so-called threshold or bucket-type photon detection, we can represent the measurement process as a Gaussian one. 
These detectors indicate either the absence of photons or the presence of at least one photon. 
Considering projection onto the vacuum state, we can define positive operator-valued measures for such detectors as a set of two operators
\begin{align}
    \hat\Pi_{\textrm{off}}=\ketbra{0}{0} \quad \textrm{and}\quad  \hat\Pi_{\textrm{on}}=\mathbb{I}-\ketbra{0}{0}.
\end{align}
Since vacuum is a Gaussian state, the projection on vacuum is a Gaussian operation, and we obtain the probability of a detection event conditioned on a given photonic state:
\begin{align}
    p_{\textrm{on}}&=\textrm{Tr}\{\hat\rho\hat\Pi_{\textrm{on}}\}=1-\textrm{Tr}\{\hat\rho\ketbra{0}{0}\}\nonumber \\
    &=1-\frac{1}{(2\pi)^N}\int d\vec{\xi}\chi_{\hat\rho}({\vec{\xi}})\chi_{\hat\Pi_{\textrm{on}}}({-\vec{\xi}}).
\end{align}
where $N$ is the number of modes and $\chi_{\hat\Pi_{\textrm{on}}}({-\vec{\xi}})$ is characteristic function of the projection operator defined as in Eq. \eqref{eq:char_func}.
Often we consider coincident detection events of a multi-mode state $\hat\rho$. 
The coincidence probability of $m$ detectors recording an event is then given by
\begin{align}
    p_{\textrm{coin}}=\textrm{Tr}\{\hat\rho & \hat\Pi^{(1)}_{\textrm{on}}\otimes \hat\Pi^{(2)}_{\textrm{on}}\otimes\dots\otimes \hat\Pi^{(m)}_{\textrm{on}} \nonumber \\
    &\otimes \mathbb{I}^{(m+1)}\otimes\dots\otimes \mathbb{I}^{(N)}\},\label{eq:coincidences}
\end{align}
which can be easily evaluated for Gaussian states using the following formula for multi-dimensional Gaussian integration
\begin{align}
   \int d\vec{x}e^{-\frac{1}{2}\vec{x}^TC\vec{x}-i\vec{d}^T\vec{x}} = \sqrt{\frac{(2 \pi)^n}{\det\left(C\right)}}\exp\left(-\frac{1}{2}\disp^TC^{-1} \disp\right) \label{eq:gauss_int}.
\end{align}

\section{Analytic expressions of figures-of-merit}
\label{sec:figuresofmerit}
We now apply the phase space formalism introduced in the previous section to model quantum teleportation of a photonic time-bin qubit.
We consider a post-selective projective Bell-state measurement (BSM) of the state $\ket{\Psi^{-}}=(\ket{el}-\ket{le})/\sqrt{2}$, where the late state $\ket{l}$ arrives after the early state $\ket{e}$, using a 50:50 beam splitter $r=t=1/\sqrt{2}$ \cite{braunstein1995measurement,bouwmeester1997,valivarthi2020}.
The time-bin qubit is in the state
$\epsilon\ket{e}+\sqrt{1-\epsilon^2}\ket{l}$, with $0\leq \epsilon \leq 1$.
The entangled qubit used to facilitate teleportation takes the form $(\ket{ee}+\ket{ll})/\sqrt{2}$.
Our model incorporates the photon fields used in Ref. \cite{valivarthi2020}, in which the qubit to be teleported is encoded into a weak coherent state $\ket{\alpha}=\mathrm{e}^{-|\alpha|^2/2}\sum_{n=0}^{\infty}(\alpha^n/\sqrt{n!})\ket{n}$, with mean photon number $|\alpha|^2$ when $|\alpha|^2\ll1$.
Two-mode squeezed vacuum $\ket{\text{TMSV}}= \sqrt{1-\mu}\sum_{n = 0}^{\infty} {\sqrt{\mu}}^n \ket{n}\ket{n}$, neglecting loss, is used in Ref. \cite{valivarthi2020} to encode the entangled state, where the kets denote the signal and the idler modes  \cite{mandel1995optical}.
This state approximates a photon pair for $\mu\ll1$ conditioned on measurement of a two-fold coincidence such that the $\ket{00}$ term is eliminated.
The mean number of pairs per signal-idler mode pair is $\mu$.
The idler mode of the two-mode squeezed vacuum is directed to the beam splitter while the signal is encoded with the teleported state at the end of the teleportation protocol.

A high-fidelity BSM, which is necessary for faithful quantum teleportation, requires photons with indistinguishable degrees of freedom \cite{braunstein1995measurement, bouwmeester1997}.
One way to characterize the distinguishability of the photons prior to teleportation is to perform HOM interference between the photons used to encode the state to be teleported and the idler mode of the entangled state \cite{hong1987}.
For HOM interference, we consider the photons to be encoded into the $\ket{e}$ bin only, neglecting all events in the $\ket{l}$ bin.

\subsection{HOM interference visibility}
\label{subsection:HOM}
\begin{figure}[h!]
    \centering
    \includegraphics[width=1\columnwidth]{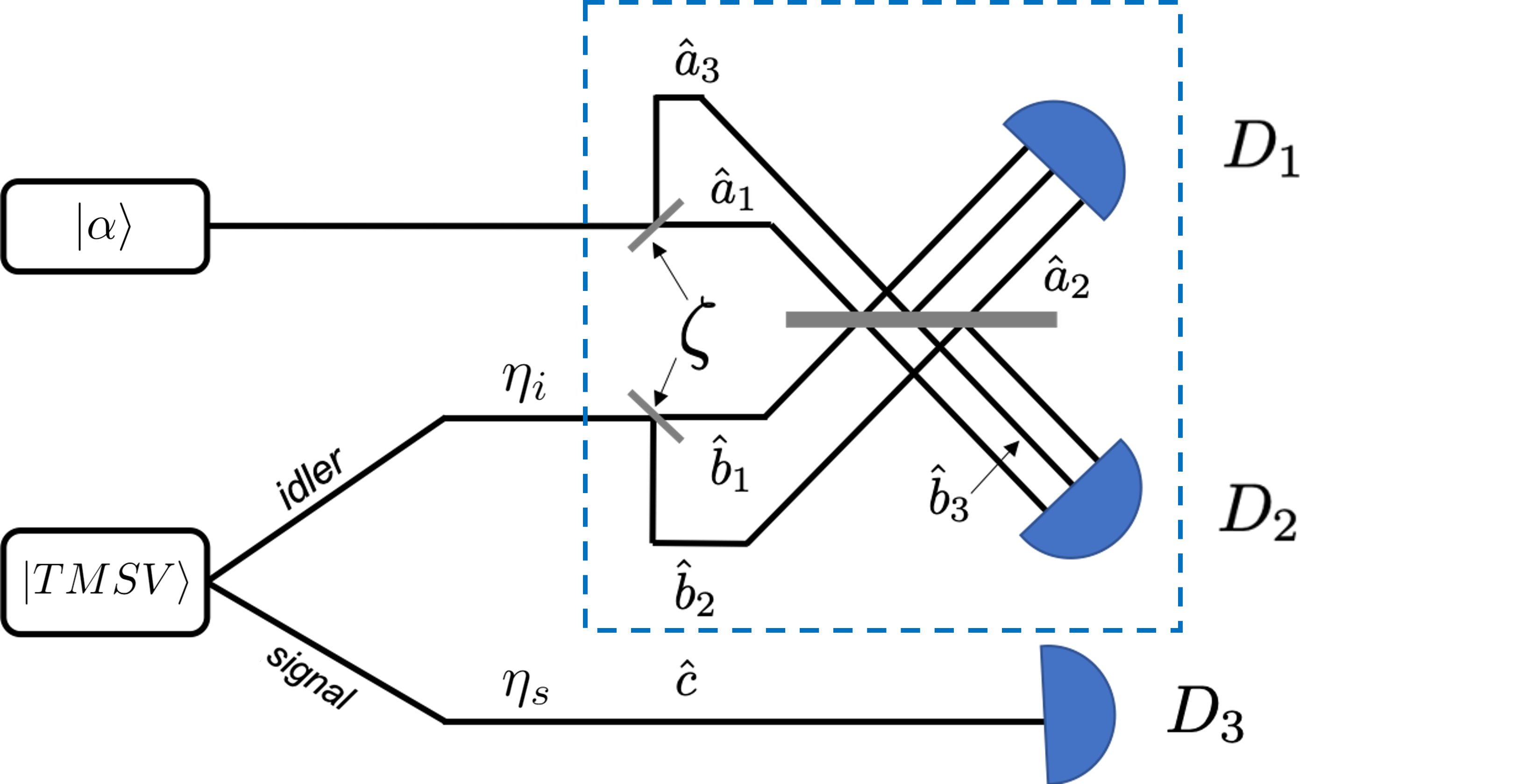}
    \caption{Model schematic for HOM interference within the context of a quantum teleportation experiment.
    The qubit to be teleported is encoded into weak coherent state $\ket{\alpha}$ whereas the entangled state is encoded into the signal and idler modes of a two-mode squeezed vacuum state $\ket{\text{TMSV}}$.
    Transmission efficiencies of the signal and idler modes are denoted by $\eta_s$ and $\eta_i$, respectively.
    HOM interference is measured by correlating detection events at $D_1$ and $D_2$ after a 50:50 beamsplitter (gray line), optionally conditioned upon detection of the signal mode at $D_3$.
    Distinguishability is modeled using virtual beamsplitters of transmittance $\zeta<1$.
    The $\hat{a}$, $\hat{b}$ and $\hat{c}$ operators refer to modes that originate from the virtual beam splitters and are used in the derivation shown in Appendix \ref{appendix:HOM}.
    Blue dashed outline is discussed in the caption of Fig. \ref{fig:teleportationmodel}.
    } \label{fig:HOMmodel}
\end{figure}

The HOM interference visibility is defined as
\begin{align}
    V_{\text{HOM}}=\frac{P_{\text{max}}-P_{\text{min}}}{P_{\text{max}}},
\end{align}
where $P_{\text{max}}$ ($P_{\text{min}}$) correspond to the probability of coincidence events between threshold detectors placed after each output of the beam splitter (cf. Fig. \ref{fig:HOMmodel}) when the photons are rendered maximally distinguishable (indistinguishable), e.g. by varying their relative polarization or time of arrival \cite{hong1986experimental,takesue20071}.
Probabilities can be converted to detection rates by multiplying them by the clock rate of the experiment.

In practice there are some additional distinguishing properties of the photons that cannot be controlled or accessed in an experiment.
We model this additional distinguishability as a finite mode mismatch $\zeta$ between the two photon fields \cite{Takeoka:2015, valivarthi2020}.
This mode mismatch is captured by a virtual beam splitter with transmittance $\zeta$ which splits each incoming photon field into indistinguishable and distinguishable parts.
Only the indistinguishable parts of both photons contribute to the interference at the actual beam splitter, whereas the distinguishable parts are combined with vacuum inputs and degrade the HOM interference visibility. 
The indistinguishability parameter $\zeta=1$ corresponds to the case in which both incoming photons are completely indistinguishable, whereas $\zeta=0$ corresponds to the case in which both photons are completely distinguishable. 

The magnitude of $V_{\text{HOM}}$ depends not only on the distinguishability of the photons but also the photon-number statistics of the fields participating in the interference \cite{knight_textbook}.
It also depends on dark counts of the detector, which we neglect for our discussion, see Sec. \ref{sec:discussion}.
Nevertheless, if the statistics are known from the experimental apparatus, then $V_{\text{HOM}}$ is an indicator of distinguishability \cite{ollivier2021hong}. 
A schematic for a HOM interference experiment is shown in Fig. \ref{fig:HOMmodel}.

There are two experimentally relevant cases to consider: the signal mode is ignored referred to as two-fold HOM interference, and when the signal mode is detected to herald the presence of ideally one idler photon, referred to as three-fold HOM interference.
For both cases we derive the two- and three-fold coincidence probabilities respectively using the characteristic function formalism assuming a weak coherent state mixing with the idler mode of two-mode squeezed vacuum as depicted in Fig. \ref{fig:HOMmodel}, with details found in Appendix \ref{appendix:HOM}.
Measurement of two-fold HOM interference is experimentally convenient, as it allows quantification of distinguishability in much less time than three-fold HOM interference.

The two- and three-fold coincidence probabilities are
 \begin{align}
 p_{\text{2-fold}}(|\alpha|^2,\mu,\zeta,\eta_i) = &1 + \frac{\exp(-|\alpha|^2)}{1+\eta_i \mu} \nonumber \\
 &-2\frac{\textrm{e}^{-\frac{|\alpha|^2}{2}\frac{[1+(1-\zeta^2)\eta_i\mu/2]}{1+\eta_i\mu/2}}}{1+\eta_i\mu/2},
\end{align}
and
\begin{align}
p_{\text{3-fold}}(|\alpha|^2,\mu,\zeta,\eta_s,\eta_i) &= \frac{\eta_s\mu}{1+\eta_s\mu} -2\frac{\textrm{e}^{-\frac{|\alpha|^2/2[1+(1-\zeta^2)\eta_i\mu/2]}{1+\eta_i\mu/2}}}{1+\eta_i\mu/2} \nonumber \\
&+\frac{\textrm{e}^{-|\alpha|^2}(1-\eta_i)\eta_s\mu}{(1+\eta_i\mu)(1+\eta_i(1-\eta_s)\mu + \eta_s\mu)} \nonumber \\
&+2\frac{\textrm{e}^{-\frac{|\alpha|^2/2[1+ (1-\zeta^2)(1-\eta_s)\eta_i\mu/2+\eta_s\mu]}{1+(1-\eta_s)\eta_i\mu/2+\eta_s\mu}}}{1+(1-\eta_s)\eta_i\mu/2+\eta_s\mu},
\end{align} 
respectively, where $\eta_s$ and $\eta_i$ are the transmission efficiencies of the signal and idler mode, respectively.
The mean photon numbers and efficiencies can be independently determined experimentally, using the methods described in Ref.~\cite{valivarthi2020}.

We now calculate the HOM interference visibility, in which $\zeta=0$ ($\zeta \leq 1$) corresponds to $P_{\text{max}}$ ($P_{\text{min}}$).
The corresponding two-fold and three-fold HOM visibilities are
\begin{multline}
V_{\text{2-HOM}}(\zeta)=1-\frac{p_{\textrm{2-fold}}(|\alpha|^2,\mu,\zeta, \eta_i)}{p_{\textrm{2-fold}}(|\alpha|^2,\mu,0, \eta_i)}\\
= -\frac{4 \left(e^{|\alpha|^2/2}-\exp\left(\frac{|\alpha|^2(2 + \eta_i \mu(1+ \zeta^2))}{2(2+\eta_i \mu)} \right)\right)(1 + \eta_i \mu )}{2 + \eta_i \mu - 4 e^{|\alpha|^2/2}(1+\eta_i \mu) + e^{|\alpha|^2}(1+\eta_i \mu)(2+\eta_i \mu)},
\label{eq:2fold_HOMvis}
\end{multline} 
and
\begin{align}
V_{\text{3-HOM}}(\zeta)&=1-\frac{p_{\textrm{3-fold}}(|\alpha|^2,\mu,\zeta,\eta_s,\eta_i)}{p_{\textrm{3-fold}}(|\alpha|^2,\mu,0,\eta_s,\eta_i)}, 
\label{eq:3fold_HOMvis}
\end{align}
respectively.
We did not expand $V_{\text{3-HOM}}$ here due to its length.

As shown in Appendix \ref{appendix:maxvisibilities}, the theoretical maximum visibilities for the two-fold and three-fold HOM visibilities are $\sqrt{2}-1\approx0.414$ and unity, respectively. 
The primary difference in maximum visibilities stems from the exclusion of $\ket{00}$ in $\ket{\text{TMSV}}$ by detection of the signal mode, that is, a single photon is heralded in the idler mode for $\mu\ll1$. 
The two-fold visibility is limited by the combination of vacuum and multi-photon events at the inputs to the beamsplitter.
Even though the twofold visibility measurement is simpler and faster, the interference fringe can be more difficult to resolve when including photon counting statistical uncertainties and potential noise.
Note that the maximum two-fold HOM visibility from interference of two coherent (thermal) fields is $1/2$ ($1/3$).

\subsection{Quantum teleportation fidelity}
\label{subsection:teleportation}
\begin{figure}[h!]
    \centering
    \includegraphics[width=1\columnwidth]{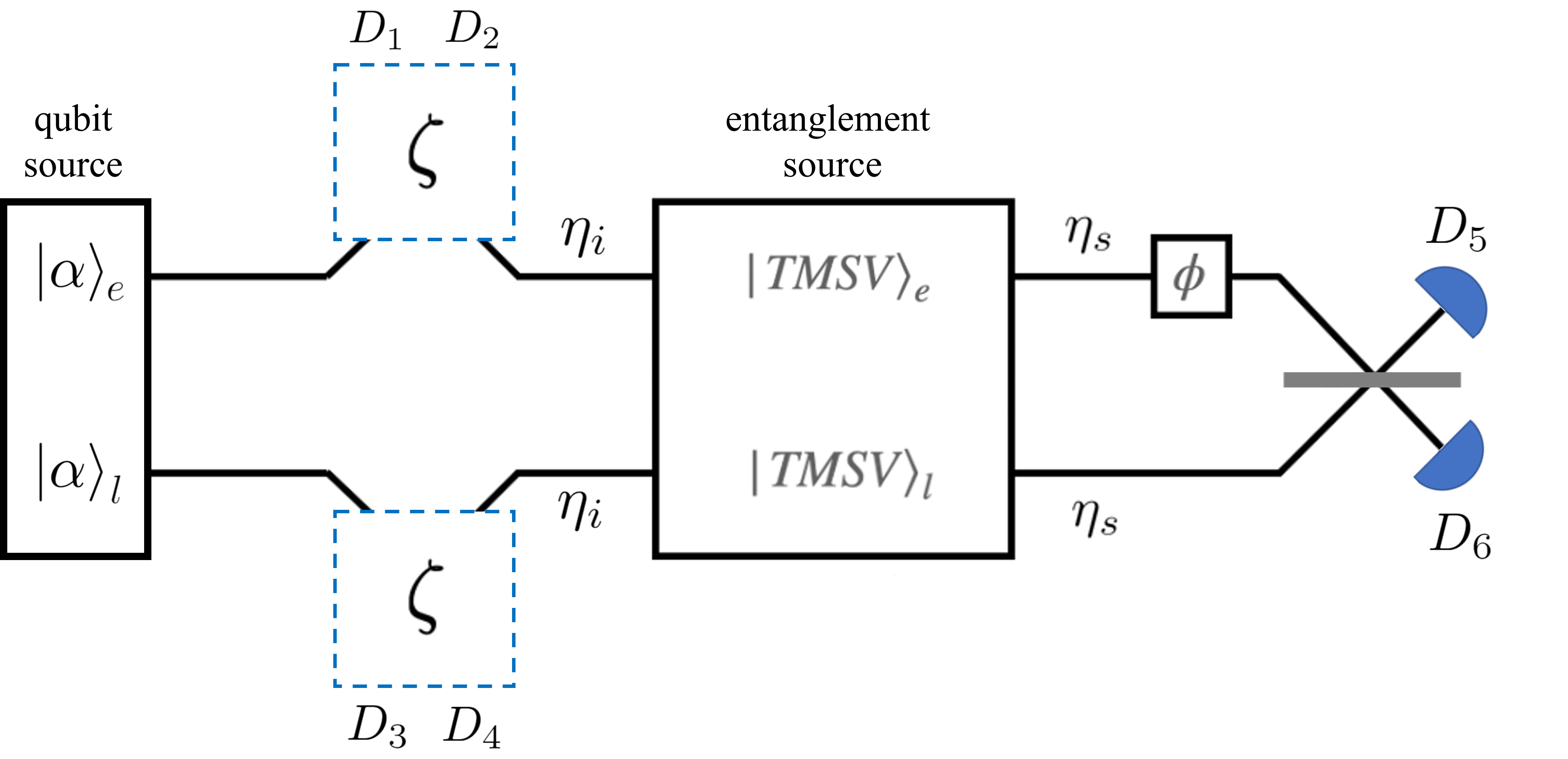}
    \caption{Model schematic of quantum teleportation.
    Each time bin, $\ket{e}$ and $\ket{l}$, is treated as a distinct spatial mode.
    As in HOM interference, the qubit to be teleported is encoded into $\ket{\alpha}$ whereas the entangled state is encoded into $\ket{\text{TMSV}}$, with relevant transmission efficiencies $\eta_s$ and $\eta_i$.
    Distinguishability of photons at the BSM is modeled using virtual beamsplitters. The indistinguishability parameter $\zeta$ outlined by the blue dashed lines corresponds to the elements enclosed by the blue dashed lines in Fig. \ref{fig:HOMmodel}.
    Projection on $\ket{\Psi^{-}}$ is indicated by coincidence detection events at $D_1$ and $D_4$ or $D_2$ and $D_3$.
    Projection of the teleported qubit onto the X-basis is modeled by a phase shift $\phi$, coherent mixing by a 50:50 beamsplitter (grey line), then photon detection at $D_5$ and $D_6$.
    Projection onto the Z-basis is modeled by removing the beamsplitter for the signal modes, that is, setting its transmittance to $t=1$, and direct detection of the photons (not shown).
    } 
    \label{fig:teleportationmodel}
\end{figure}

We extend the HOM interference model to that for quantum teleportation by including the $\ket{l}$ time-bin in the analysis.
As depicted in Fig. \ref{fig:teleportationmodel}, we model the $\ket{e}$ and $\ket{l}$ bins as independent spatial modes, with interference for the BSM taking place at the beamsplitters ascribed for the $\ket{e}$ modes and that for the $\ket{l}$ modes.
Distinguishability is again modeled using virtual beamsplitters like the HOM interference model.
Projection on $\ket{\Psi^{-}}$ is indicated by a specific coincidence detection event between a photon in the $\ket{e}$ and $\ket{l}$ bins.
Specifically this corresponds to a photon being detected in $\ket{e}$ ($\ket{l}$) in one detector and $\ket{l}$ ($\ket{e}$) in the other \cite{valivarthi2014efficient}.
Measurement of the teleported qubit in the Z-basis, that is, the $\ket{e}$ or $\ket{l}$ mode, is modeled by detection of the photon in a distinct spatial mode.
Measurement in the X-basis, that is, in the state $(\ket{e}+e^{i\phi}\ket{l})/\sqrt{2}$, is modeled by combining each spatial signal mode on a 50:50 beamsplitter after introducing a relative phase $\phi$, then detecting each photon in a distinct spatial mode. 
In other words, the measurement basis is rotated, as facilitated by phase-sensitive interference.

Using the characteristic function formalism, as shown in Appendix~\ref{appendix:teleportation}, we derive the teleportation fidelity for the X-basis states using
\begin{align}
    F=\frac{P_{D_1D_4D_6}(\phi)}{P_{D_1D_4D_6}(\phi) +P_{D_1D_4D_5}(\phi)},
    \label{eq:teleportationX}
\end{align}
where $P_{D_1D_4D_6}$ ($P_{D_1D_4D_5}$) is the coincidence detection probability for a $\ket{\Psi^{-}}$ projection and measurement of the teleported qubit in the intended (orthogonal) state \cite{nielsen2010quantum}.
The argument $\phi$ indicates that the measurement basis is oriented to the intended state of the teleported qubit such that $P_{D_1D_4D_6}$ ($P_{D_1D_4D_5}$) is maximized (minimized).
Note that the detection event corresponding to $P_{D_2D_3}$ also corresponds to a projection onto $\ket{\Psi^{-}}$.
These events are treated like those corresponding to $P_{D_1D_4}$ due to symmetry, consistent with properties of the beamsplitters and detectors used for the BSM in Ref. \cite{valivarthi2020}.

Considering teleportation of Z-basis states, the corresponding teleportation fidelity takes the same form as Eq. \ref{eq:teleportationX}, but with different and $\phi$-independent underlying expressions for $P_{D_1D_4D_6}$, and similarly $P_{D_1D_4D_5}$, because the beamsplitter used for measurement of X-basis states is removed.
See Appendix~\ref{appendix:teleportation} for details.
The maximum theoretical X- and Z-basis teleportation fidelities are one.

\section{Fit of model with experimental results of Ref.~\cite{valivarthi2020}}
\label{sec:fit}

Using the expressions for two- and three-fold HOM interference visibilities and teleportation fidelity, we fit our model to data from Ref.~\cite{valivarthi2020} to reveal the imperfections in the experiment and validate our model.
We consider only teleportation of X-basis states in this section since they sufficiently capture the behavior of Z-basis states and they are sensitive to $\zeta$.

\begin{figure}[h!]
    \centering
    \includegraphics[width=1.0\columnwidth]{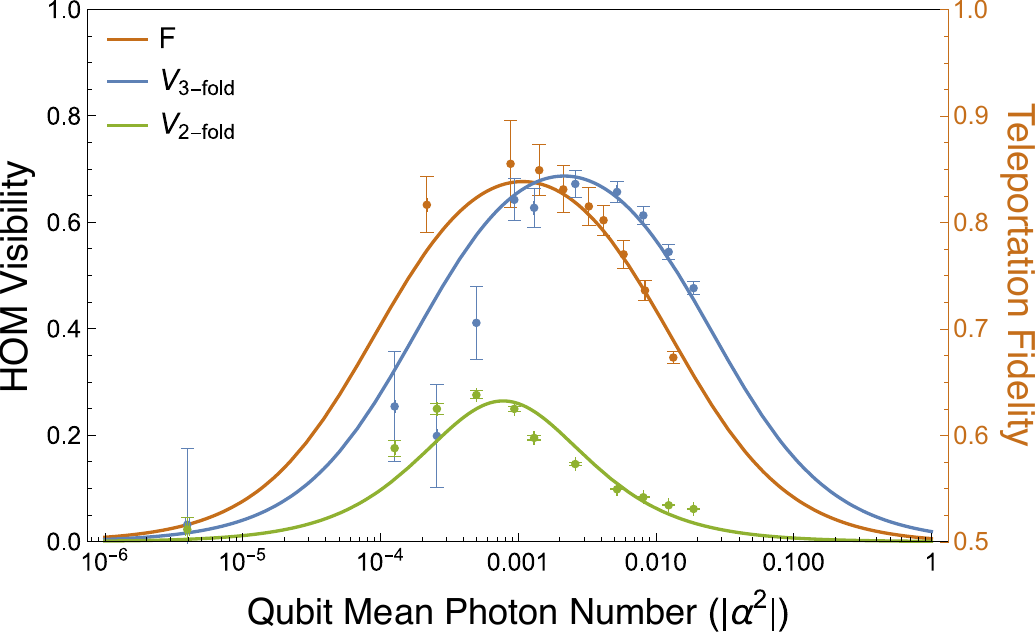}
    \caption{Two- and three-fold HOM interference visibilities ($V_\text{2-fold}$, green and $V_\text{3-fold}$, blue) and quantum teleportation fidelity ($F$, red) of X-basis states for varied qubit mean photon number $|\alpha|^2$.
    The model (lines) is fit to, and agrees with, the experimental data of Ref. \cite{valivarthi2020} (points).
    The mean photon number is shown on a log scale to provide a simple representation of the model.}
    \label{fig:threefolddataplots}
\end{figure}

First, we plot the experimentally measured two- and three-fold HOM interference visibilities as well as teleportation fidelity for varied mean photon number $|\alpha|^2$ of the weak coherent state used to encode the qubit.
This is shown in Fig. \ref{fig:threefolddataplots}.
We choose to probe our figures-of-merit against $|\alpha|^2$ because of the experimental ease to vary this parameter and, as discussed in Ref. \cite{valivarthi2020}, the use of $|\alpha|^2$ for preparing decoy states.
We consider $|\alpha|^2<1$ since this is most relevant regime in teleportation experiments.
The maximum measured two- and three-fold HOM visibilities and teleportation fidelity are $0.28\pm0.01$, $0.67\pm0.03$, and $0.86\pm 0.04$, respectively, due to distinguishability and undesired multi-photon events from $\ket{\alpha}$ and in the idler mode \cite{zhong2013nonlocal}.
Note that the three-fold HOM interference visibility as predicted by the teleportation fidelity $2F-1\sim72\%$ is consistent \cite{werner1989}.

The teleportation fidelity and HOM visibilities decrease to 0.5 and 0, respectively, for very large or small values of $|\alpha|^2=0$.
As $|\alpha|^2$ increases from zero, there are more events in which a single photon from $\ket{\alpha}$  and a single photon in the idler mode contribute to HOM interference and the BSM, and thus the visibility rises.
Since the probability of two photons in $\ket{\alpha}$ also grows, there will be two-fold detection events that correspond to vacuum in the idler, due to idler field statistics or loss, which reduces the visibility.
Therefore, the trade-off between interference events originating from single- or multi-photon states arriving at the beamsplitter leads to the visibility reaching a maximum and decreasing for higher $|\alpha|^2$.
The specific value of $|\alpha|^2$ that corresponds to the maximum visibility is also conditioned on whether a two- or three-photon detection experiment is performed, as well as the values of $\mu$, $\eta_i$, and $\eta_s$.
This interpretation as well as the curve shapes and positions for two- and three-fold experiments are further discussed in Appendix \ref{appendix:maxvisibilities} and Sec. \ref{sec:analysis}.

For the two-fold data, the maximum occurs at $|\alpha|^2= 5.0\times10^{-4}$ while the three-fold visibility and teleportation fidelity are maximized around $|\alpha|^2 = 2.6\times10^{-3}$ and $|\alpha|^2 = 8.8\times10^{-4}$, respectively.
The value of $|\alpha|^2$ corresponding to maximum two-fold HOM visibility is less than that for the maximum three-fold visibility because three-fold events are conditioned on detecting at least one photon in the signal mode. 
As a result, there are fewer vacuum events in the idler mode in the three-fold case, and thus $|\alpha|^2$ can be increased, i.e. the probability of $n=1$ and $n=2$ photon events can be increased, to reach maximum visibility.
The three-fold HOM visibility data maximizes at a $|\alpha|^2$ that is a factor of two higher than the teleportation data because $|\alpha|^2$ is defined per qubit, which corresponds to the mean photon number in two temporal modes.

Next, we proceed to fit Eqs. \ref{eq:2fold_HOMvis}, \ref{eq:3fold_HOMvis}, and \ref{eq:teleportationX} to each of the relevant data sets according to the procedure discussed in Appendix \ref{appendix:fitting}.
We ascribe a different mode mismatch parameter, $\zeta_{2}$ and $\zeta_{3}$, for the two- and three-fold detection experiments, respectively. 
This originates from the multi-mode $\ket{\text{TMSV}}$ used in Ref. \cite{valivarthi2020}.
Although spectral filtering of the signal and idler modes was employed in Ref. \cite{valivarthi2020}, detection of the signal can further filter the spectrum of the idler field due to residual non-zero frequency entanglement \cite{davis2022, Kim_delayedchoice}.
For the same reason, we also ascribe different idler mode efficiencies, $\eta_{i2}$ and $\eta_{i3}$, for the two- and three-fold detection experiments, respectively, as heralding a photon in the signal mode effectively removes frequency modes from the idler, which manifests as additional inefficiency in the idler mode.

The three-fold HOM and teleportation data is fit together using a shared $\zeta_{3}$, and with the following independently measured parameters from Ref. \cite{valivarthi2020} held constant: $\eta_{i3}=1.2\times10^{-2}$, $\eta_s=4.5\times10^{-3}$, and $\mu=8.0\times10^{-3}$.
Note that these parameters were measured in Ref. \cite{valivarthi2020} using coincidence detection of the filtered $\ket{\text{TMSV}}$ with $|\alpha|^2=0$.
The two-fold HOM data is fitted separately with only $\mu=8.0\times10^{-3}$ held constant.
The fits reveal $\zeta_2 = 0.80 \pm 0.04$, $\eta_{i2}=(6.9\pm1.2)\times10^{-2}$ and $\zeta_3 = 0.90 \pm 0.02$.
The fitted curves are plotted in Fig. \ref{fig:threefolddataplots}, and are in good agreement with the measured data.
Furthermore, $\zeta_3$ matches that fitted in Ref. \cite{valivarthi2020}.
The fits clearly reveal that heralding removes additional frequency components to improve indistinguishability close to unity, which underpins the high teleportation fidelity observed in Ref. \cite{valivarthi2020}. 
The fits also yield $\eta_{i2}>\eta_{i3}$ as expected.

Notice the curves take on the form of a log-normal distribution, which owes to the Poission distribution of number states in $\ket{\alpha}$.
When plotted on a linear scale, as shown in Ref. \cite{valivarthi2020}, the long tail in the distribution for $|\alpha|^2>>10^{-3}$ can be interpreted as the trade-off between interference produced by the $n=2$ term in $\ket{\alpha}$ with vacuum in the idler, and from single photons in $\ket{\alpha}$ and the idler.

It is convenient to infer the value of $|\alpha|^2$ that will maximize $V_{\text{HOM}}$ for a given experimental setup.
Thus, in Appendix \ref{appendix:HOM_calculus} we differentiate Eqs. \ref{eq:2fold_HOMvis} and \ref{eq:3fold_HOMvis}, finding $|\alpha|^2$ of $7.8 \times 10^{-4}$ and $2.2 \times 10^{-3}$, for two- and three-fold HOM interference experiments, respectively, 
consistent with the data shown in Fig. \ref{fig:threefolddataplots}.

\section{Prediction of figures-of-merit under varying experimental conditions}
\label{sec:analysis}
We now employ our analytical model to further interpret the experimental imperfections in Ref. \cite{valivarthi2020} and to predict the outcomes of future experiments under varying experimental imperfections: indistinguishability, transmission efficiencies, and mean photon numbers of $\ket{\alpha}$ and $\ket{\text{TMSV}}$ states.
For simplicity, from now on we assume a  $\ket{\text{TMSV}}$ such that heralding does not vary the indistinguishability or the path loss for the idler mode.
We also assume the idler path efficiency to be identical for two- and three-fold detection experiments.

\subsection{Indistinguishability}
\label{subsection:indistinguishability}
\begin{figure*}[ht!]
    \centering
    \includegraphics[width = \textwidth]{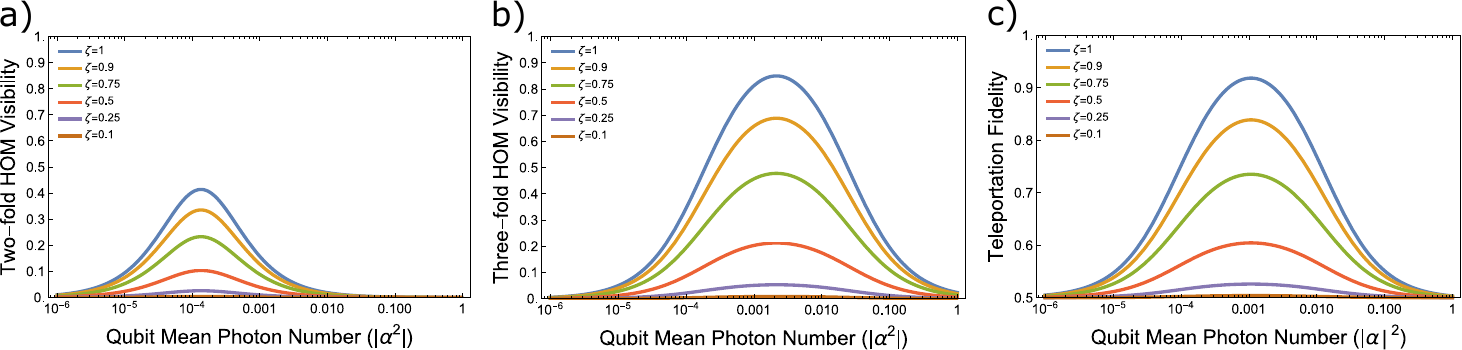}
    \caption{
    Model of a) two- and b) three-fold HOM interference visibilities as well as c) teleportation fidelity of X-basis states for varied $|\alpha|^2$ and different magnitudes of indistinguishability $\zeta$ between the interfering photons.
    The curves assume the transmission efficiencies and $\mu$ from the three-fold detection experiments of Ref. \cite{valivarthi2020}.
    }
    \label{fig:varyzeta}
\end{figure*}

To determine the role of distinguishability, in Fig. \ref{fig:varyzeta} we plot two- and three-fold HOM interference visibilities as well as X-basis teleportation fidelity as a function of $\zeta$ under the experimental conditions of three-fold detection from Ref. \cite{valivarthi2020}: $\eta_i=\eta_{i3}=1.2\times10^{-2}$, $\eta_s=4.5\times10^{-3}$, and $\mu=8.0\times10^{-3}$.
Our model predicts a simple vertical scaling of the curves. 
The maximum visibilities and fidelity still occur at the same $|\alpha|^2$, but with increased maxima.
It is also experimentally convenient that the optimum value of $|\alpha|^2$ is independent of distinguishability.
The curves retain the log-normal behavior, supporting our interpretation that the curve shape is due to mean photon number mismatch.
Note the two-fold curve has shifted to a lower central value of $|\alpha|^2$ compared to that in Fig. \ref{fig:threefolddataplots} due to the reduction of idler path efficiency in the model henceforth compared to that in the experimental results.

The model predicts maximum two- and three-fold HOM visibilities as well as X-basis teleportation fidelity to increase to $\sqrt{2}-1$, 0.85 and 0.92, respectively, for completely indistinguishable fields.
The two-photon visibility curve reaches the theoretical maximum because idler mode inefficiency does not vary the number statistics of the idler mode, and as discussed in Appendix \ref{appendix:maxvisibilities}, $|\alpha|^2=\sqrt{2}\mu$ maximizes the visibility in the regime of low mean photon numbers we consider here.
The maximum three-fold HOM visibility and teleportation fidelity does not reach unity due to multi-photon components from $\ket{\alpha}$ and non-unit transmission of the idler mode.
Yet, reasonably high teleportation fidelity can be achieved even with significant path loss ($\sim1\%$) provided $\mu$ is kept low.
Note that if the 0.98 fidelity of the Z-basis states from Ref. \cite{valivarthi2020} is included, the total average fidelity will reach 0.94.

Without loss of generality, we assume complete indistinguishability $\zeta=1$ for all remaining plots in the manuscript to probe the dependence of the other experimental imperfections. 
Notice that the X-basis teleportation fidelity curves follow the same dependence as the three-fold HOM interference visibility curves.
Thus, to avoid , we move all of the relevant teleportation curves to Appendix \ref{appendix:teleportationcurves}.
Note that the factor of two shift in $|\alpha|^2$ between three-fold HOM interference visibility and fidelity is retained for all curves.

\subsection{Transmission efficiencies}
\label{subsection:efficiencies}

We compare signal and idler mode transmission efficiencies of Ref. \cite{valivarthi2020}, i.e. those plotted in Fig. \ref{fig:threefolddataplots} with $\eta_i=1.2\times10^{-2}$ and $\eta_s=4.5\times10^{-3}$, to those of unit efficiency.
We assume $\mu=8.0\times10^{-3}$ as before and plot the two- and three-fold detection curves under four different configurations: 
\begin{enumerate}
  \item[(i)] $\eta_i=\eta_s=1$
  \item[(ii)] $\eta_i=1.2\times10^{-2}$ and $\eta_s=4.5\times10^{-3}$ in Ref. \cite{valivarthi2020} 
  \item[(iii)] $\eta_i=1$ and $\eta_s=4.5\times10^{-3}$
  \item[(iv)] $\eta_i=1.2\times10^{-2}$ and $\eta_s=1$
  \end{enumerate}

\begin{figure}[h!]
    \centering
    \includegraphics[width = \columnwidth]{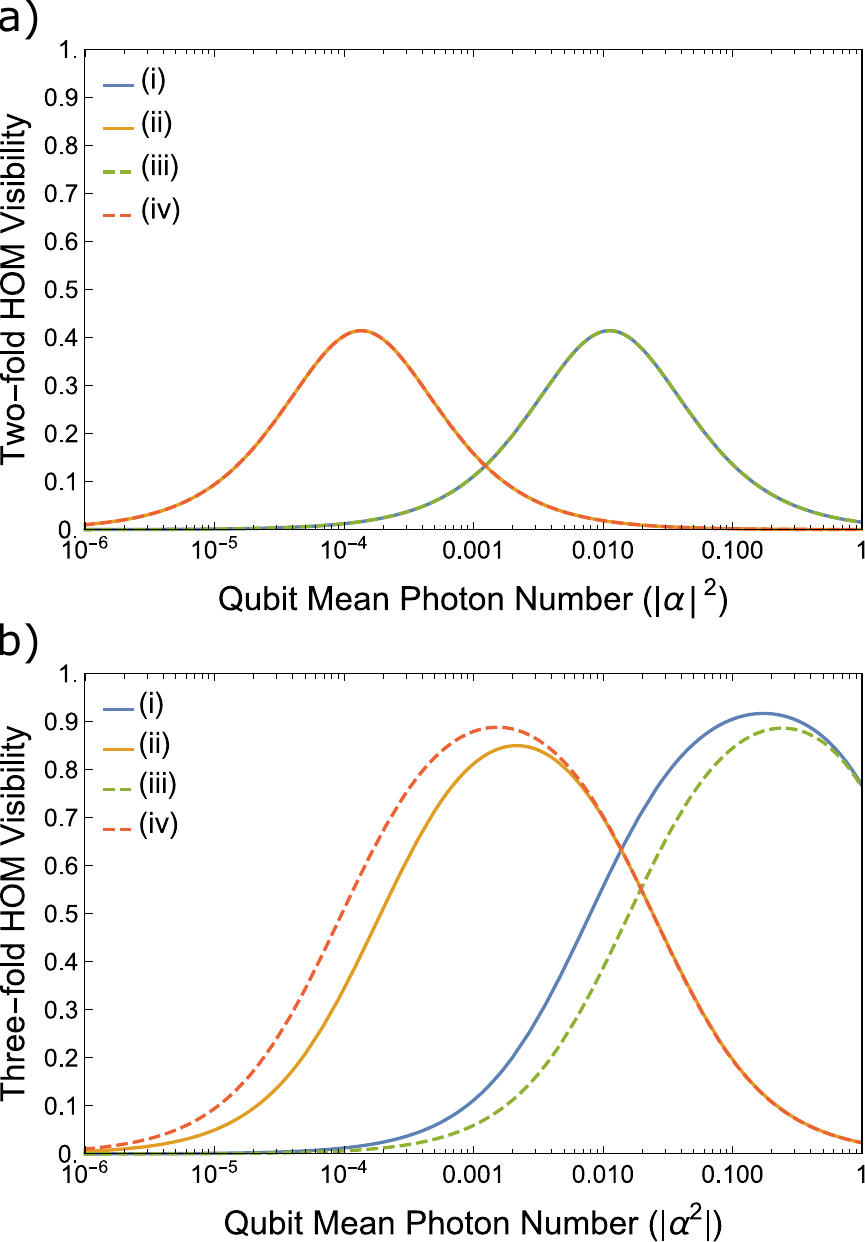}
    \caption{Model of a) two- and b) three-fold HOM interference visibilities for varied $|\alpha|^2$ under conditions of varied signal and idler transmission efficiencies (i-iv) in blue, orange, green, and red, respectively, as described in the main text, assuming $\mu=8.0\times10^{-3}$ and complete indistinguishability $\zeta=1$.   
    For the two-fold HOM curves, configurations (i) and (ii) are equivalent to (iii) and (iv), respectively.
    }
    \label{fig:varyeta}
\end{figure}

For the two-fold HOM visibility curves in Fig. \ref{fig:varyeta}, we find a reduction of idler transmission efficiency from unity (cases (i) and (iii)) to $\eta_i$ (cases (ii) and (iv)) retains the curve profile, but shifts it to be centered around a value of $|\alpha|^2$ that is a factor of $\eta_i$ lower.
As discussed in Appendix \ref{appendix:maxvisibilities}, the visibility is maximized when $|\alpha|^2=\sqrt{2}\mu$, that is, the mean photon numbers of $|\alpha|^2$ and the idler mode match to a scaling factor.
Since the mean photon number of the idler scales proportional to the idler transmission efficiency, $|\alpha|^2$ must be reduced by the same factor to maximize visibility.
The curve shape does not change with $\eta_i$ because $\mu<<1$.
The maximum visibility is saturated to its theoretical maximum of $\sqrt{2}-1$ in the low mean photon number regime here, as discussed in Sec. \ref{subsection:indistinguishability}. 

For the three-fold HOM visibility curves, the unity transmission case (i) shifts the distribution from Ref. \cite{valivarthi2020} (ii) to higher $|\alpha|^2$ to better match the effective higher mean photon number in the idler mode.
The maximum visibility increases for (i) because detection of a photon in the signal mode is always accompanied by a photon in the idler mode i.e. zero vacuum components in the idler mode, that is, the photon heralding efficiency \cite{klyshko1980} is unity. 
However, the maximum visibility is not unity, primarily due to multi-photon components from $\ket{\alpha}$ and, to some extent, the idler mode.

For case (iii), the added loss in the signal mode compared to (i) removes some single pair events in the experiment.
Thus, multi-photon events from $\ket{\text{TMSV}}$ are relatively more likely to be detected by the threshold detectors.
By way of heralding, this leads to a relative increase in multi-pair events from $\ket{\text{TMSV}}$, reducing the maximum visibility from (i) while effectively increasing the mean photon number in the idler mode, thereby forcing an increase in $|\alpha|^2$.
The width of the curve is also reduced from the lower $|\alpha|^2$ edge, which is consistent with the increase in multi-photon events from $\ket{\text{TMSV}}$.
To understand this, consider $|\alpha|^2$ being lowered from the value that maximizes the visibility.
In this case, the increased multi-photon events in the idler mode reduce the visibility more strongly than the case in which non-zero vacuum is present in the idler.
This narrowing becomes even more pronounced as $\mu$ is increased, as discussed in Sec. \ref{subsection:mu}.
The curve narrowing effect is not observed in the two-fold visibility curves because heralding, and hence the effect of signal path inefficiency, strongly changes the number distribution in the idler mode.

\begin{figure}[h!]
    \centering
    \includegraphics[width=\columnwidth]{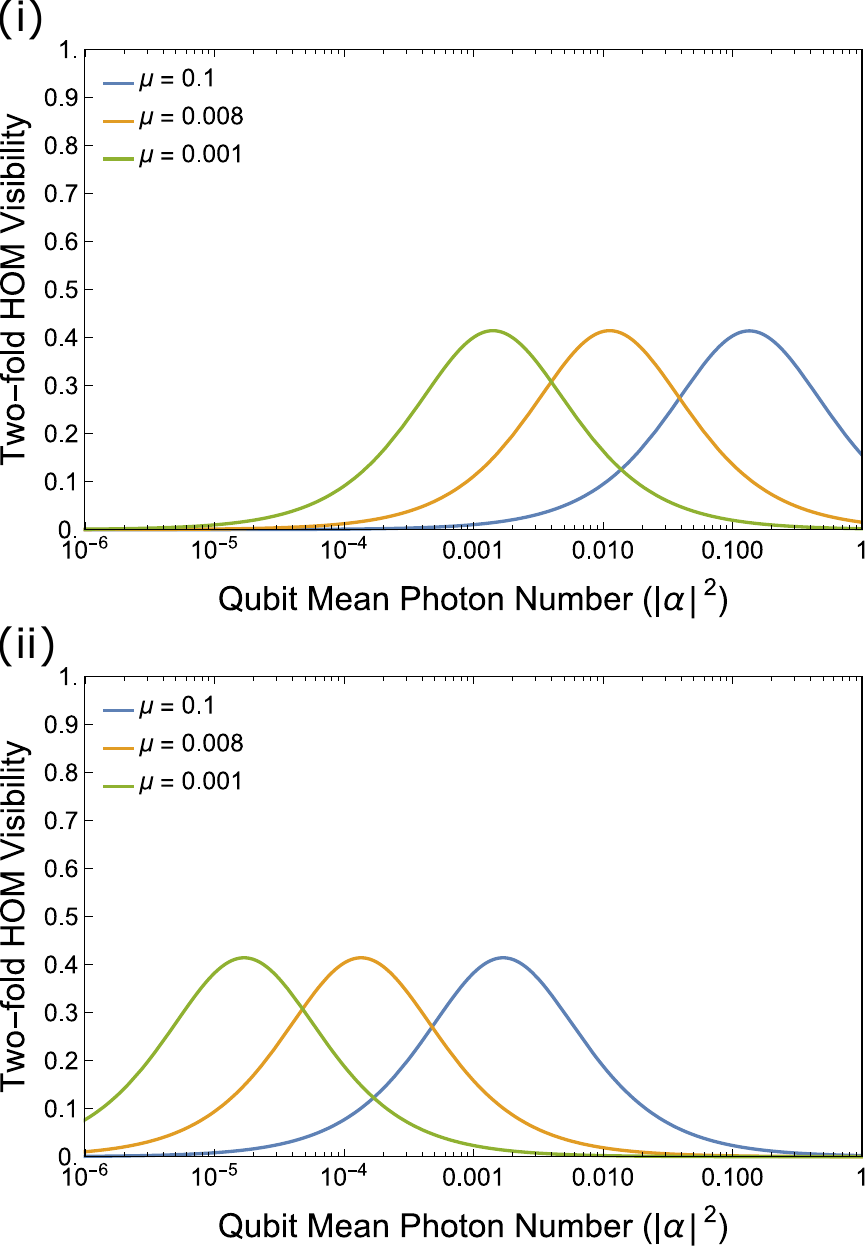}
    \caption{Model of two-fold HOM interference visibilities for varied $|\alpha|^2$ and $\mu<10^{-2}$, under varied signal and idler transmission efficiencies, cases (i) and (ii), which are equivalent to (iii) and (iv), respectively, assuming complete indistinguishability $\zeta=1$.
    }
    \label{fig:varymu_2fold}
\end{figure}

For case (iv), the shift of the curve to lowered $|\alpha|^2$ compared to case (ii) is similar to that when comparing case (i) to (iii).
It is due to the relative decreased contribution of multi-photon detection events in the idler mode, which must be matched by $|\alpha|^2$.
The maximum visibility is limited by the non-unit heralding efficiency; when a single photon is detected in the signal mode, it may not be present at the beamsplitter and multi-photon events originating from $\ket{\alpha}$ have a relatively higher probability of contributing to the visibility.
This is also why maximum visibility is not as high as case (i), but higher than the rest of the curves which have low signal efficiency and for which heralding cannot benefit as much.
Akin to that observed when comparing curves (i) and (iii), the width of curve (iv) increases from the lower $|\alpha|^2$ edge relative to case (ii).
This is also consistent with the presence of more heralded single photons, and fewer multi-photon events from heralding.
The high $|\alpha|^2$ edge of curves in cases (ii) and (iv), and (i) and (iii), converge with increased $|\alpha|^2$ because the Poisson- distributed multi-photon components from $\ket{\alpha}$ reduce the visibility more strongly than the heralded components of $\ket{\text{TMSV}}$ in this regime.

\subsection{Mean photon number of $\ket{\text{TMSV}}$}
\label{subsection:mu}
Here we compare $\mu=10^{-1}$, $\mu=8.0\times10^{-3}$ from Ref. \cite{valivarthi2020}, and $\mu=10^{-3}$ under cases (i)-(iv).
Two-fold HOM interference visibilities with varied $|\alpha|^2$ are plotted in Fig. \ref{fig:varymu_2fold} under these scenarios.

For case (i)/(iii) and (ii)/(iv) we observe the same behavior as in Fig. \ref{fig:varyeta}a.
The curve profile shifts to lowered $|\alpha|^2$ to ensure the mean photon number of $|\alpha|^2$ matches that in the idler mode when either the idler transmission efficiency or, equivalently, mean photon number of $\ket{\text{TMSV}}$ is reduced.
The shift in $|\alpha|^2$ matches the reduction in $\mu$, and the overall shift of the three curves when the idler path efficiency is reduced is the same as discussed in Sec. \ref{subsection:efficiencies}.
Furthermore, the maximum visibility can reach its theoretical maximum under these conditions, as discussed in Secs. \ref{subsection:indistinguishability} and \ref{subsection:efficiencies}.

\begin{figure*}[ht!]
   \centering
   \includegraphics[width=\textwidth]{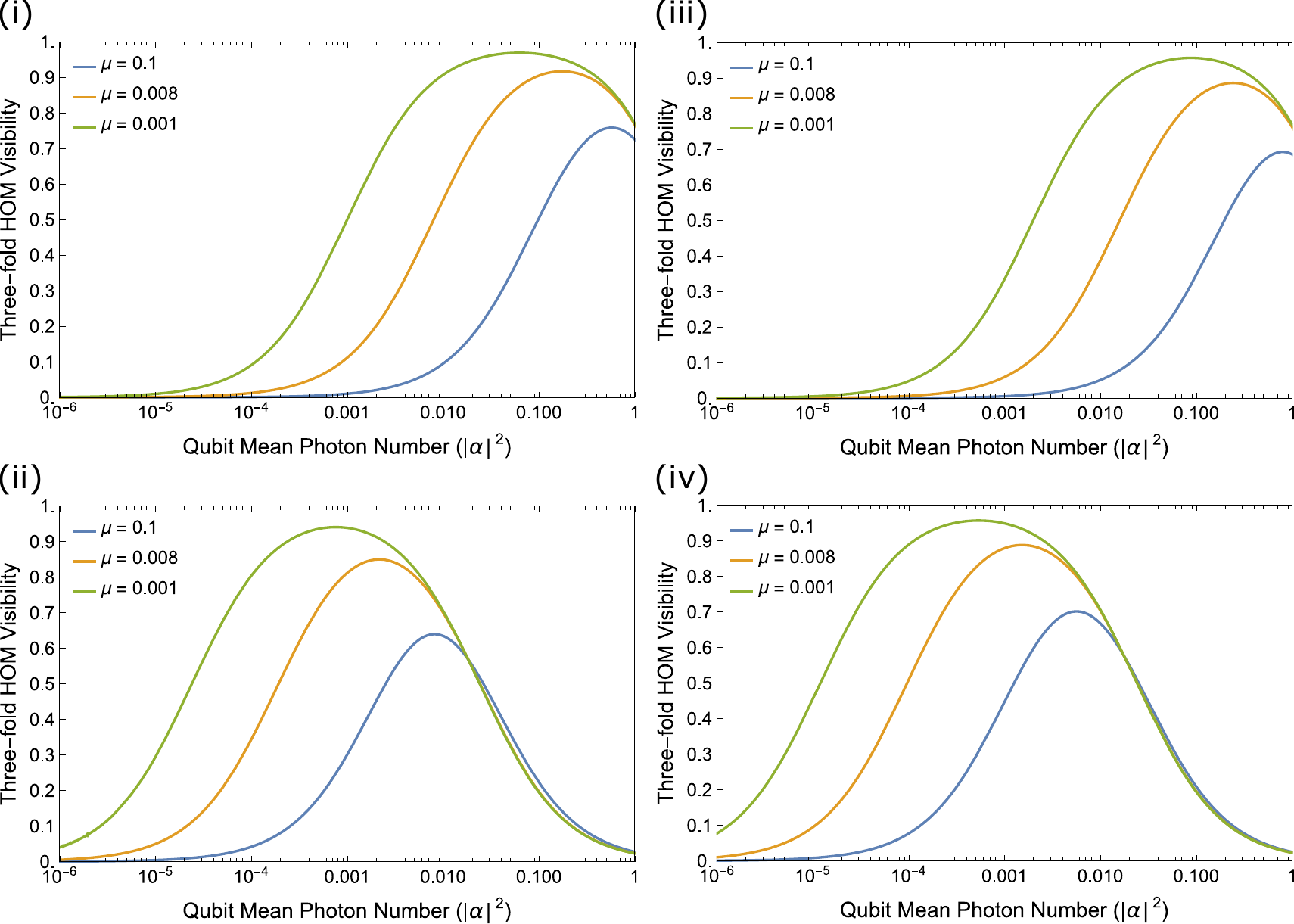}
    \caption{Model of three-fold HOM interference visibilities for varied $|\alpha|^2$ and $\mu<10^{-2}$, under varied signal and idler transmission efficiencies, cases (i)-(iv), assuming complete indistinguishability $\zeta=1$.
    }
    \label{fig:varymu_3fold}
\end{figure*}

The three-fold HOM interference visibility is plotted under these scenarios (i)-(iv) in Fig. \ref{fig:varymu_3fold}.
We discuss case (i) first.
To give a point of reference, the curve corresponding to $\mu=8.0\times10^{-3}$ matches the case (i) curve shown in Fig. \ref{fig:varyeta}b.
Interestingly, the family of curves appear to have a similar behavior to those in Fig. \ref{fig:varyeta}b, except now with $\mu$ varied instead of signal path efficiency.
An increase of $\mu$ increases the number of multi-photon states that are heralded which both lowers the maximum visibility and requires a higher $|\alpha|^2$ to match, similar to that discussed for the curves in Fig. \ref{fig:varyeta}b.
The width of the curve is also reduced from the lower $|\alpha|^2$ edge as $\mu$ increases, which is again consistent with the increase in multi-photon events from $\ket{\text{TMSV}}$.
For high $\mu$, as $|\alpha|^2$ is lowered from the value corresponding to maximum visibility, the curve falls more sharply than others because there are more multi-photon terms in $\ket{\text{TMSV}}$ and  $\ket{\alpha}$, and a slight mismatch in mean photon numbers will lead to higher order terms contributing a larger reduction in visibility.
For the case in which $\mu$ is smaller, there are fewer multi-photon terms from $\ket{\text{TMSV}}$ in the idler mode, and also $|\alpha|^2$, which leads to a broader peak on the low $|\alpha|^2$ edge.
A slight mismatch of single photon probabilities is not accompanied by strong multi-photon effects in this scenario.
As in Fig. \ref{fig:varyeta}b, the high $|\alpha|^2$ edge of curves converge with increased $|\alpha|^2$ because the Poisson-distributed multi-photon components from $\ket{\alpha}$ dominate in this regime relative to conditions for maximum visibility.

The curves for case (ii), which corresponds to the non-unit path efficiencies in Ref. \cite{valivarthi2020}, have remarkably similar shapes to case (i) with a few differences.
The shift of the distributions to lowered $|\alpha|^2$ is again due to the idler path loss.
The maximum visibilities are lowered because the signal loss reduces the number of single photon events in the idler path that contribute to interference, and the idler loss further decreases this number, which leads to multi-photon terms from both $\ket{\text{TMSV}}$ and $\ket{\alpha}$ contributing.
This also explains why the curve widths are also reduced compared to case (i) even though the curves are centered around lower values of $|\alpha|^2$; the multi-photon terms quickly dominate when mean photon numbers are not matched.
Note that the narrowing of the distribution from case (i) to case (ii) is also slightly observed in Fig. \ref{fig:varyeta}.
Observe that the $\mu=0.1$ curve edge slightly extends over the others for high $|\alpha|^2$ because $\ket{\text{TMSV}}$ contributes more terms in the idler mode in this regime compared to the others.

The family of curves for case (iii), which corresponds to non-ideal signal path efficiency, differs from case (i) much in the same way that these cases differ in Fig. \ref{fig:varyeta}.
That is, the curves take on a similar form as (i), but with lower maximum visibilities, maxima that are shifted to higher $|\alpha|^2$, and narrower curve widths compared to those in (i).
The shape and offset of the curves follows reasons previously discussed, which owe to increased multi-photon events in the idler mode.
For case (iv), corresponding to non-ideal idler path efficiency, differs from case (ii) again much in the same way that these cases differ in Fig. \ref{fig:varyeta}.
The unit signal efficiency now increases the visibilities, shifts the curve to lower $|\alpha|^2$, and broadens the curve widths compared to those in (ii).
The shape and offset again follow reasons previously discussed.
The broadening is particularly pronounced for low $\mu$, which also requires low $|\alpha|^2$, and thus very few multi-photon events contribute, and hence are less effected by idler path loss.

\section{Discussion} \label{sec:discussion} Our analytical expressions for realistic photonic quantum teleportation experiments with time-bin qubits are valuable for guiding the design and optimization of future experiments. Achieving transmission efficiencies or indistinguishability beyond 99\% in typical photonics experiments requires significant effort \cite{alexander2024manufacturable}. Our modeling quantifies the improvements provided by such efforts under different experimental configurations and indicates the effort needed to meet minimum acceptable standards for various applications, such as quantum communication. Additionally, our analytical expressions allow for predicting experimental outcomes using independently measured parameters, including indistinguishability, which can be estimated through mode measurements like laser linewidth or cavity resonance profiles.

By quantifying our figures-of-merit against the log of the mean photon number $|\alpha|^2$ of an input weak coherent state $\ket{\alpha}$, we find a simple log-normal distribution that aids in interpreting and utilizing our analytical expressions. The curves simplify the role of indistinguishability to a simple scaling of visibility or teleportation fidelity. The two-fold HOM interference visibility curves are the quickest to interpret and, along with their rapid measurement compared to three-fold HOM interference, are valuable for prototyping setups. For teleportation, a low mean photon number of photon pairs $\mu$ strongly mitigates path inefficiency and relaxes the precision needed for the value of $|\alpha|^2$ that maximizes visibility. Conversely, a relatively high $\mu \sim 0.1$ significantly reduces teleportation fidelity but not below the classical bound of $2/3$ for our parameter range. Even with path efficiencies of $\sim 1\%$, the reduction is not severe, with signal path inefficiency impacting visibilities to a lesser extent. This is unsurprising given the number of successful quantum networking experiments using lossy setups or links that require extended data collection periods. However, higher mean numbers of pairs necessitate careful calibration of $|\alpha|^2$ to maximize visibility. The data in Fig. \ref{fig:threefolddataplots} shows that calibration of $|\alpha|^2$ in Ref. \cite{valivarthi2020} was challenging for $|\alpha|^2 << 10^{-3}$. Nonetheless, we find good agreement between our analytical expressions and the measurement data from Ref. \cite{valivarthi2020}, which spans almost four decades of $|\alpha|^2$.

Although our modeling captures all relevant behavior in the experiment of Ref. \cite{valivarthi2020} and can be applied to other photonic quantum teleportation experiments, future work could include more detailed modeling of the multi-mode nature of $\ket{\text{TMSV}}$. This involves incorporating the effects of pump bandwidth and frequency filtering, as done in Ref. \cite{davis2022}. The Schmidt decomposition of $\ket{\text{TMSV}}$ approximates the number of modes, and the filter acts as a mode-selective beamsplitter \cite{zielnicki2018joint}. This would relate differences in indistinguishability and loss in the idler path for two- and three-fold detection to specific apparatus configurations. It is also straightforward to incorporate noise or detector dark counts into the modeling. Moreover, our methods can be extended to non-Gaussian measurements, including photon number resolved detection, which can improve heralding efficiencies of single photons \cite{davis2022}.

Our modeling applies to different discrete-variable encodings beyond time-bin and readily extends to more complex experiments such as entanglement swapping or GHZ-state generation. The use of Gaussian states and transformations also extends to other experiments using bosonic modes, such as atomic ensembles or other parametric interactions like electro-optics or opto-mechanics, and their relevant applications in communications, computing, and sensing. Although an analysis based on "photon counting" in the Fock basis could have been used to analyze the outcomes in Ref. \cite{valivarthi2020}, we believe that our presented analysis provides an intuitive picture of the underlying physics with a compact, experimentally realistic, and "universal" methodology that can be easily extended to other experimental operating regimes, such as using squeezing \cite{gurses2024free}.

\begin{acknowledgments}
The authors acknowledge support from the AQT Intelligent Quantum Networks and Technologies (AQT/INQNET, established in 2017) research program. S.D. is partially supported by the Brinson Foundation. N.L., M.S., S.D., and D.R.M. acknowledge partial support from the U.S. Department of Energy (DoE), Office of Science, High Energy Physics, QuantISED program (Award No. DE-SC0019219 QCCFP) and Basic Energy Sciences (Award No. DE-SC0020376, HEADS-QON for transduction-related work). K.T. acknowledges the Samuel P. and Frances Krown SURF fellowship. R.V., M.S., and S.D. are partially supported by the DoE, ASCR, Advanced Quantum Networks for Scientific Discovery (AQNET-SD). R.V. and M.S. are also partially supported by the DoE's ASCR QUANT-NET project. The core of this work was conducted between 2020 and 2023 following our publication in PRX Quantum 1, 020317, 2020, with the manuscript prepared in 2024. M.S. is grateful to the Caltech SURF program for partial support of undergraduate research. Three of the authors worked on this project alongside their SURF research, and two have since moved on to graduate school.

\end{acknowledgments}

\onecolumngrid
\appendix 
\section{Analytical Derivations of Expressions}

\subsection{HOM Interference Visibility}
\label{appendix:HOM}
We employ the characteristic function formalism described in Sec. \ref{section:characteristicfunction} considering the setup shown in Fig. \ref{fig:teleportationmodel}.
For this derivation, we use 7$\times$7 block matrices with 2$\times$2 sub-matrices, with each sub-matrix representing correlations between different optical modes.
The first column of the block matrix represents the coherent state mode; the third and fifth columns represent vacuum inputs at the virtual beamsplitters with transmission $\zeta^2$ to account for the mode indistiguishability, the second and sixth columns represent the vacuum inputs at the 50:50 beamsplitter that are mixed with the distinguishable parts of the modes; and the fourth and the seventh columns represent the idler and signal modes of the TMSV state. 
We first describe the overall state of the system after transmission losses, given by the block covariance matrix $$\covar = \begin{pmatrix} 
\bf{I_{2x2}} & 0 & 0 & 0 & 0 & 0 & 0 \\
0 & \bf{I_{2x2}}  & 0 & 0 & 0 & 0 & 0 \\
0 & 0 & \bf{I_{2x2}}  & 0 & 0 & 0 & 0 \\
0 & 0 & 0 & (1 + 2 \eta_i \mu) \bf{I_{2x2}} & 0 & 0 & 2 \sqrt{\eta_i \eta_s \mu (1 + \mu)} \sigma_3  \\
0 & 0 & 0 & 0 &\bf{I_{2x2}} & 0 & 0 \\
0 & 0 & 0 & 0 & 0 & \bf{I_{2x2}} & 0 \\
0 & 0 & 0 & 2 \sqrt{\eta_i \eta_s \mu (1 + \mu)} \sigma_3 & 0 & 0 & (1 + 2 \eta_s \mu) \bf{I_{2x2}}\\
\end{pmatrix},$$ where $\sigma_3 = \begin{pmatrix}1 & 0 \\ 0 & -1  \end{pmatrix}$.
The displacement vector is $\disp = \sqrt{2} \begin{pmatrix} \Re(\alpha) & \Im(\alpha) & 0 & ... & 0\end{pmatrix}^T$, with $\alpha$ already accounting for loss in the coherent state channel. 
From here, we apply the mismatch matrix $$\begin{pmatrix} \sqrt{\zeta} \  \bf{I_{2x2}} & 0 &  \sqrt{1-\zeta} \ \bf{Z} & 0 & 0 & 0 & 0 \\ 0 & \bf{I_{2x2}} & 0 & 0 & 0 & 0 & 0 \\ \sqrt{1-\zeta} \ \bf{Z} & 0 & \sqrt{\zeta} \  \bf{I_{2x2}} & 0 & 0 & 0 & 0 \\ 0 & 0 & 0 & \sqrt{\zeta} \  \bf{I_{2x2}} & \sqrt{1-\zeta} \  \bf{Z} & 0 & 0 \\ 0 & 0 & 0 &  \sqrt{1-\zeta} \  \bf{Z} & \sqrt{\zeta} \  \bf{I_{2x2}} & 0 & 0 \\ 0 & 0 & 0 & 0 & 0 & \bf{I_{2x2}} & 0 \\  0 & 0 & 0 & 0 & 0 & 0 & \bf{I_{2x2}} \end{pmatrix}$$ and the beam splitting matrix $$\frac{1}{\sqrt{2}} \begin{pmatrix}  \bf{I_{2x2}} & 0 & 0 & \bf{Z} & 0 & 0 & 0 \\ 0 &\bf{I_{2x2}} & 0 & 0 & \bf{Z} & 0 & 0 \\ 0&0 &\bf{I_{2x2}} & 0 & 0 & \bf{Z} & 0 \\ \bf{Z} & 0 & 0 & \bf{I_{2x2}} & 0 & 0 & 0 \\ 0 & \bf{Z} & 0 & 0 & \bf{I_{2x2}} & 0 & 0 \\ 0 & 0 & \bf{Z} & 0 & 0 & \bf{I_{2x2}} & 0 \\  0 & 0 & 0 & 0 & 0 & 0 & \sqrt{2} \ \bf{I_{2x2}} \end{pmatrix},$$ where $\bf{Z} = \begin{pmatrix} 0 & -1 \\ 1 & 0 \end{pmatrix}$.
This now allows calculation of the twofold coincidence probability:
\begin{align}p_{\text{2-fold}} = \textrm{Tr}\big\{\dense' \big( 
& 1 - \mathbb{I}_{b_1, b_2, b_3, c} \otimes  \ketbra{0}{0}^{\otimes 3}_{a_1, a_2, a_3} \nonumber\\
&- \mathbb{I}_{a_1, a_2, a_3, c} \otimes  \ketbra{0}{0}^{\otimes 3}_{b_1, b_2, b_3} \nonumber\\
&+ \mathbb{I}_{c} \otimes  \ketbra{0}{0}^{\otimes 6}_{a_1, a_2, a_3, b_1, b_2, b_3} \big) \big\},\label{eq:2foldappendix} \end{align} 
and the threefold coincidence probability:
\begin{align}p_{\text{3-fold}} = \textrm{Tr}\big\{\dense' \big( 
& 1 - \mathbb{I}_{a_1, a_2, a_3, b_1, b_2, b_3} \otimes \ketbra{0}{0}_{c} \nonumber \\ &- \mathbb{I}_{b_1, b_2, b_3, c} \otimes  \ketbra{0}{0}^{\otimes 3}_{a_1, a_2, a_3} \nonumber \\
&- \mathbb{I}_{a_1, a_2, a_3, c} \otimes  \ketbra{0}{0}^{\otimes 3}_{b_1, b_2, b_3} \nonumber\\ &+ \mathbb{I}_{b_1, b_2, b_3} \otimes  \ketbra{0}{0}^{\otimes 4}_{a_1, a_2, a_3, c} \nonumber\\
&+ \mathbb{I}_{a_1, a_2, a_3} \otimes  \ketbra{0}{0}^{\otimes 4}_{b_1, b_2, b_3, c}\nonumber\\
&+ \mathbb{I}_{c} \otimes  \ketbra{0}{0}^{\otimes 6}_{a_1, a_2, a_3, b_1, b_2, b_3} \nonumber\\
&- \ketbra{0}{0}^{\otimes 7}_{a_1, a_2, a_3, b_1, b_2, b_3, c} \big) \big\},\label{eq:3foldappendix} \end{align}
where the subscripts $a_i, b_i,$ and $c$, with $i\in\{1,2,3\}$, represent the coherent state, idler, and  signal modes respectively. 
Note that the subscripts of the identity matrices indicate which modes are traced out for a given calculation. 
Using Eq. \ref{eq:gauss_int}, we calculate $$\textrm{Tr}\{ \dense (\mathbb{I}_{{x_1, ..., x_m}} \otimes \ketbra{0}{0}^{\otimes n}_{y_1, ..., y_n}\} = \frac{2^n}{\sqrt{\det\left(\covar_{y_1, ..., y_n} + \mathbb{I}\right)}}\exp\left(-\disp_{y_1, ..., y_n}^T(\covar_{y_1, ..., y_n} + \mathbb{I})^{-1}\disp_{y_1, ..., y_n} \right),$$ where $x_i$ represents the modes traced out and $y_i$ represents the remaining modes. 
We now use this expression to calculate each of the terms in Eqs. \ref{eq:2foldappendix} and \ref{eq:3foldappendix}, yielding \begin{align*} p_{\text{2-fold}}(|\alpha|^2,\mu,\zeta,\eta_i) = &1 + \frac{\exp(-|\alpha|^2)}{1+\eta_i \mu} \\
&- 4\frac{\exp\left(-|\alpha|^2 + \frac{|\alpha|^2 (2 + (1 + \zeta^2)\eta_i \mu)}{4 + 2 \eta_i \mu} \right)}{2 + \eta_i \mu},
\end{align*}
and
\begin{align*}
p_{\text{3-fold}}(|\alpha|^2,\mu,\zeta,\eta_s,\eta_i) &= \frac{\eta_s\mu}{1+\eta_s\mu} -2\frac{\textrm{e}^{-\frac{|\alpha|^2/2[1+(1-\zeta^2)\eta_i\mu/2]}{1+\eta_i\mu/2}}}{1+\eta_i\mu/2} \nonumber \\
&+\frac{\textrm{e}^{-|\alpha|^2}(1-\eta_i)\eta_s\mu}{(1+\eta_i\mu)(1+\eta_s(1-\eta_s)\mu + \eta_s\mu)} \nonumber \\
&+2\frac{\textrm{e}^{-\frac{|\alpha|^2/2[1+ (1-\zeta^2)(1-\eta_s)\eta_i\mu/2+\eta_s\mu]}{1+(1-\eta_s)\eta_i\mu/2+\eta_s\mu}}}{1+(1-\eta_s)\eta_i\mu/2+\eta_s\mu}.
\end{align*}

\subsection{Teleportation Fidelity}
\label{appendix:teleportation}

We now consider the setup of Fig. \ref{fig:teleportationmodel}. 
Similar to that in Sec. \ref{appendix:HOM}, we use 14$\times$14 block matrices with 2$\times$2 sub-matrices, with each sub-matrix representing correlations between different optical modes. 
The first and the eight columns of the block matrices represent the early and late coherent state modes; the fourth, and eleventh columns represent the early and late idler modes; the seventh and fourteenth columns represents the early and late signal mode; and the rest represents the vacuum inputs at the virtual and the 50:50 beamsplitters.  
Again, the block covariance matrix denotes the state of the system after losses:
\setcounter{MaxMatrixCols}{20}
$$ \bf{I_{6x6}} \oplus \begin{pmatrix}
( 1 + 2 \eta_s \mu) \bf{I_{2x2}}& 0& 0& 0& 0& 0& 0& 0& 0 & 2 \sqrt{\eta_s \eta_i \mu (1 + \mu)} \sigma_3 & 0\\ 
 0& \bf{I_{2x2}}& 0& 0& 0& 0& 0& 0& 0& 0& 0\\
0& 0& \bf{I_{2x2}}& 0& 0& 0& 0& 0& 0& 0& 0\\
0& 0& 0& \bf{I_{2x2}}& 0& 0& 0& 0& 0& 0& 0\\ 
0& 0& 0& 0& \bf{I_{2x2}}& 0& 0& 0& 0& 0& 0\\
 0& 0& 0& 0& 0& \bf{I_{2x2}}& 0& 0& 0& 0& 0\\
0& 0& 0& 0& 0& 0& ( 1 + 2 \eta_s \mu) \bf{I_{2x2}}& 0& 0& 0& 2 \sqrt{\eta_s \eta_i \mu (1 + \mu)} \sigma_3\\ 
0& 0& 0& 0& 0& 0& 0& \bf{I_{2x2}}& 0& 0& 0\\
 0& 0& 0& 0& 0& 0& 0& 0& \bf{I_{2x2}}& 0& 0\\
 2 \sqrt{\eta_s \eta_i \mu (1 + \mu)} \sigma_3& 0& 0& 0& 0& 0& 0& 0& 0& ( 1 + 2 \eta_i \mu) \bf{I_{2x2}} & 0\\
0& 0& 0& 0& 0& 0& 2 \sqrt{\eta_s \eta_i \mu (1 + \mu)} \sigma_3 & 0& 0& 0& (1 + 2 \eta_i \mu) \bf{I_{2x2}}.
\end{pmatrix}$$ 
The displacement vector is $\disp = \sqrt{2} \begin{pmatrix} \Re(\alpha) & \Im(\alpha) & 0 & 0 & 0 & 0 & 0 & 0 & 0 & 0 & 0 & 0 & \Re(\alpha) & \Im(\alpha) & 0 & 0 & ... & 0 \end{pmatrix}^T$, and $\alpha$ again already takes into account loss.
The mismatch matrix $$\begin{pmatrix} \sqrt{\zeta} \  \bf{I_{2x2}} & 0 &  \sqrt{1-\zeta} \ \bf{Z} & 0 & 0 & 0  \\ 0 & \bf{I_{2x2}} & 0 & 0 & 0 & 0  \\ \sqrt{1-\zeta} \ \bf{Z} & 0 & \sqrt{\zeta} \  \bf{I_{2x2}} & 0 & 0 & 0  \\ 0 & 0 & 0 & \sqrt{\zeta} \  \bf{I_{2x2}} & \sqrt{1-\zeta} \  \bf{Z} & 0  \\ 0 & 0 & 0 &  \sqrt{1-\zeta} \  \bf{Z} & \sqrt{\zeta} \  \bf{I_{2x2}} & 0 \\ 0 & 0 & 0 & 0 & 0 & \bf{I_{2x2}} \end{pmatrix} \oplus \begin{pmatrix} \sqrt{\zeta} \  \bf{I_{2x2}} & 0 &  \sqrt{1-\zeta} \ \bf{Z} & 0 & 0 & 0  \\ 0 & \bf{I_{2x2}} & 0 & 0 & 0 & 0  \\ \sqrt{1-\zeta} \ \bf{Z} & 0 & \sqrt{\zeta} \  \bf{I_{2x2}} & 0 & 0 & 0  \\ 0 & 0 & 0 & \sqrt{\zeta} \  \bf{I_{2x2}} & \sqrt{1-\zeta} \  \bf{Z} & 0  \\ 0 & 0 & 0 &  \sqrt{1-\zeta} \  \bf{Z} & \sqrt{\zeta} \  \bf{I_{2x2}} & 0 \\ 0 & 0 & 0 & 0 & 0 & \bf{I_{2x2}} \end{pmatrix} \oplus \bf{I_{4x4}}$$ is applied, and so is the beam splitting matrix $$\frac{1}{\sqrt{2}} \begin{pmatrix}  \bf{I_{2x2}} & 0 & 0 & \bf{Z} & 0 & 0 \\ 0 &\bf{I_{2x2}} & 0 & 0 & \bf{Z} & 0 \\ 0&0 &\bf{I_{2x2}} & 0 & 0 & \bf{Z} \\ \bf{Z} & 0 & 0 & \bf{I_{2x2}} & 0 & 0 \\ 0 & \bf{Z} & 0 & 0 & \bf{I_{2x2}} & 0 \\ 0 & 0 & \bf{Z} & 0 & 0 & \bf{I_{2x2}} \\ \end{pmatrix} \oplus \frac{1}{\sqrt{2}} \begin{pmatrix}  \bf{I_{2x2}} & 0 & 0 & \bf{Z} & 0 & 0 \\ 0 &\bf{I_{2x2}} & 0 & 0 & \bf{Z} & 0 \\ 0&0 &\bf{I_{2x2}} & 0 & 0 & \bf{Z} \\ \bf{Z} & 0 & 0 & \bf{I_{2x2}} & 0 & 0 \\ 0 & \bf{Z} & 0 & 0 & \bf{I_{2x2}} & 0 \\ 0 & 0 & \bf{Z} & 0 & 0 & \bf{I_{2x2}} \\ \end{pmatrix} \oplus \bf{I_{4x4}}.$$ 
The above result is in the Z-basis. 
For the X-basis, we apply the phase shift matrix to the early signal mode $$\bf{I_{12x12}} \oplus \begin{pmatrix} \cos(\phi) & \sin(\phi) \\  -\sin(\phi) & \cos(\phi) \end{pmatrix} \oplus \bf{I_{14x14}}$$ and then interfere the early and late signal mode at a 50:50 beamsplitter described by the matrix 
$$\bf{I_{12x12}} \oplus \begin{pmatrix} \frac{1}{\sqrt{2}}\bf{I_{2x2}} & 0 & 0 & 0 & 0 & 0 & 0 & \frac{1}{\sqrt{2}}\bf{Z} \\ 0 & \bf{I_{2x2}} & 0 & 0 & 0 & 0 & 0 & 0\\ 0 & 0 & \bf{I_{2x2}} & 0 & 0 & 0 & 0 & 0 \\ 0 & 0 & 0 & \bf{I_{2x2}} & 0 & 0 & 0 & 0\\ 0 & 0 & 0 & 0 & \bf{I_{2x2}} & 0 & 0 & 0 \\ 0 & 0 & 0 & 0 & 0 & \bf{I_{2x2}} & 0 & 0 \\0 & 0 & 0 & 0 & 0 & 0 & \bf{I_{2x2}} & 0 \\ \frac{1}{\sqrt{2}}\bf{Z} & 0 & 0 & 0 & 0 & 0 & 0 & \frac{1}{\sqrt{2}}\bf{I_{2x2}} \end{pmatrix}$$
before the detection.
In both the X- and Z-basis, we calculate the relevant threefold detection probabilities as follows: \begin{align}P_{D_1 D_4 D_6} = \textrm{Tr}\big\{\dense' \big( 
 1 &- \mathbb{I}_{a_l, b_e, b_l, c_e, c_l} \otimes \ketbra{0}{0}^{\otimes 3}_{a_e} - \mathbb{I}_{a_e, a_l, b_e, c_e, c_l}\otimes \ketbra{0}{0}^{\otimes 3}_{b_l} \nonumber\\ &- \mathbb{I}_{a_e, a_l, b_e, b_l, c_e} \otimes  \ketbra{0}{0}_{c_l}   + \mathbb{I}_{a_l, b_e, b_l, c_e} \otimes  \ketbra{0}{0}^{\otimes 4}_{a_e, c_l} \nonumber\\ &+ \mathbb{I}_{a_e, a_l, b_e, c_e} \otimes  \ketbra{0}{0}^{\otimes 4}_{b_l, c_l}
+ \mathbb{I}_{a_l, b_e, c_e, c_l} \otimes  \ketbra{0}{0}^{\otimes 6}_{a_e, b_l} \nonumber\\ &- \mathbb{I}_{a_l, b_e, c_e} \otimes \ketbra{0}{0}^{\otimes 7}_{a_e, b_l, c_l} \big) \big\}, \label{eq:3foldteleportationappendix} \end{align} where the subscripts $a, b,$ and $c$ represent the coherent state, idler, and signal modes respectively, and the subscripts $e$ and $l$ represent the early or late bin, respectively.
Again, the subscripts of the identity matrix indicate which modes are traced out for a given calculation. 
Similarly as before, we use Eq.~\ref{eq:gauss_int} to calculate each of the terms in Eq. \ref{eq:3foldteleportationappendix} to yield analytical expressions of the probabilities.

\section{Maximum theoretical HOM interference visibilities}
\label{appendix:maxvisibilities}
\begin{figure}[!tbp]
  \centering
  \subfloat[]{\includegraphics[width=0.49\textwidth]{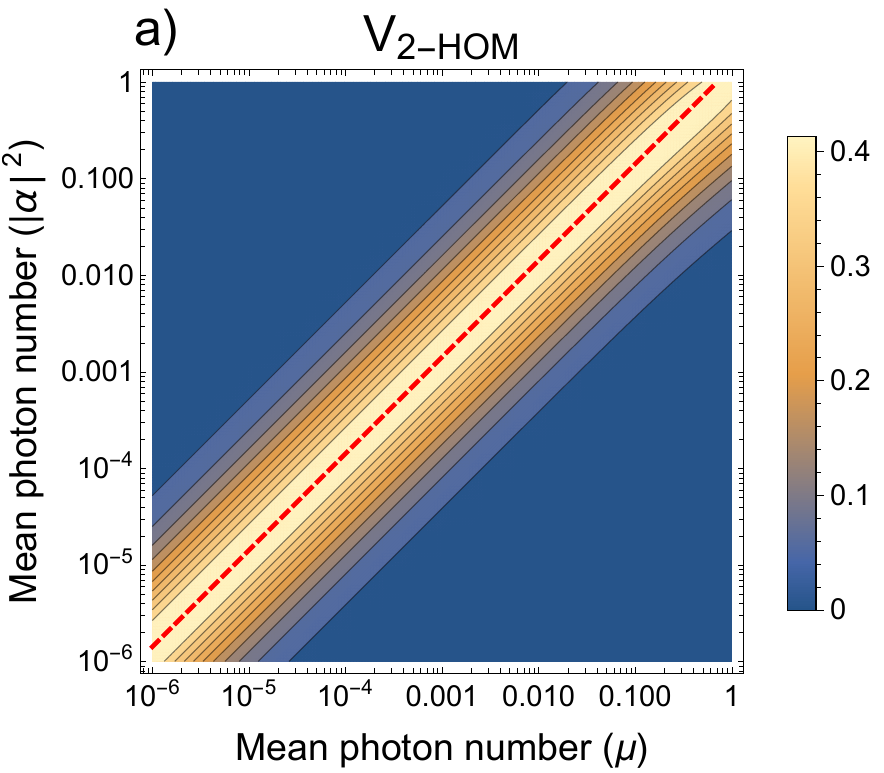}\label{fig:f1}}
  \hfill
  \subfloat[]{\includegraphics[width=0.49\textwidth]{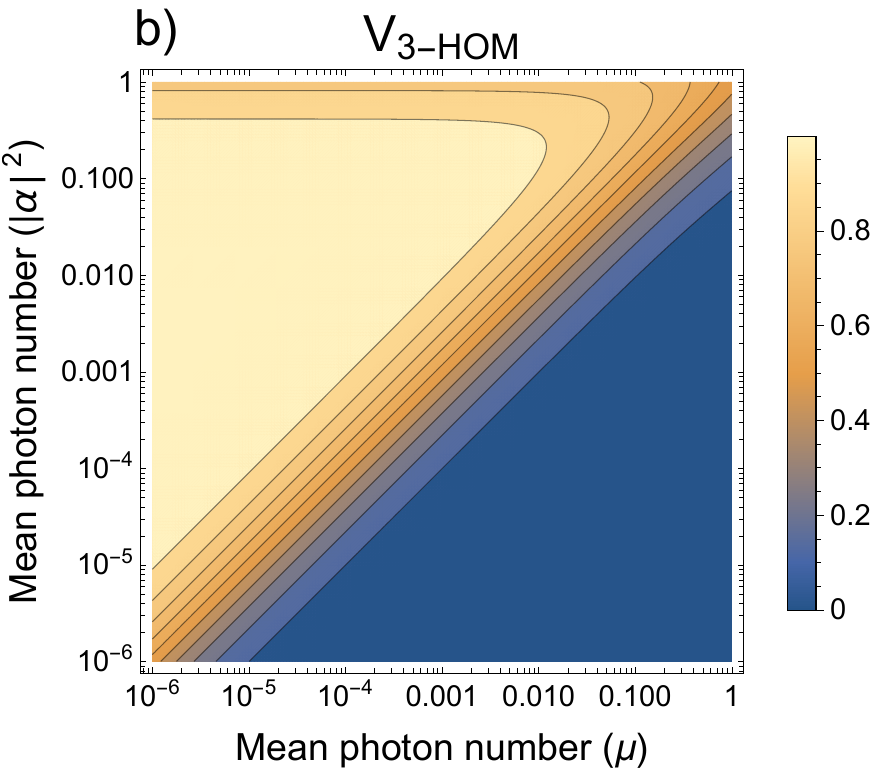}\label{fig:f2}}
  \caption{Dependence of a) two-fold and b) three-fold HOM interference visibilities for varied mean photon numbers of the coherent state ($|\alpha|^2$) and TMSV ($\mu$) assuming unity path efficiencies ($\eta_{i}, \eta_{s} = 1$) and photon indistinguishability ($\zeta=1$). The red dashed line in a) corresponds to $|\alpha|^2 = \sqrt{2}\mu$, which maximizes the visibility for $|\alpha|^2,\mu<<1$.}  \label{fig:vis_heatmaps}
\end{figure}

We plot the maximum two- and three-fold interference HOM interference visibilities using Eqs. \ref{eq:2fold_HOMvis} and \ref{eq:3fold_HOMvis}.
Complete indistinguishability $\zeta=1$ as well as perfect transmission $\eta_s=\eta_i=1$ is assumed.
The visibilities with varied $|\alpha|^2$ and $\mu$ are shown in Fig. \ref{fig:vis_heatmaps}, finding maximum two- and three-fold visibilities of $\sqrt{2}-1$ and unity, respectively.

The two-fold plot features a maximum along a symmetric diagonal for all $|\alpha|^2<<1$ and $\mu<<1$.
The maximum corresponds to the condition $|\alpha|^2/\mu=\sqrt{2}$, which is not equal to one due to the differences in number distributions.
This condition effectively corresponds to matching of the mean photon numbers of the Poisson and thermal distributions, `striking a balance" between the contributions of single photons interfering compared to $n=2$ terms interfering with vacuum.
Thus, the $\sqrt{2}$ acts to ensure that the balance is struck between the different field statistics.
This is different than the case of identical field statistics, in which the maximum corresponds to an exact matching of mean photon numbers.
Note the slight deviation in diagonal symmetry as $|\alpha|$ and $\mu$ approach 1; the balancing offered by $\sqrt{2}$ cannot hold because interference between single and $n=2$ states begin to contribute to interference.
Indeed the maximum visibility is not unity due to the non-zero contribution of $n=2$ terms interfering with vacuum.

Owing to heralding, the three-fold plot has a plateau-like topography that extends the range of optimized visibility. 
A range of $|\alpha|^2<<1$ will maximize the visibility to approach unity because the measurement is conditioned on three-fold detection and heralding will always guarantee a single photon in the idler mode when $\mu<<1$.
Effectively, this regime renders the visibility independent of the probability of generating a photon in $\ket{\alpha}$.
The threshold at $|\alpha|^2\sim1$ is predominantly due to $n=2$ events from $\ket{\alpha}$ interfering with heralded single photons in the idler path, thereby reducing the maximum visibility.
A steep diagonal threshold to the plateau is also present under conditions of $|\alpha|^2<<1$, similar to that of the ridge in the two-fold plot.
In this region, as $\mu$ is increased and approaches $|\alpha|^2$, the relative probability of heralding a multi-photon term increases, which decreases the visibility, and leads to the threshold topography along the diagonal.
The condition $|\alpha|^2/\mu=\sqrt{2}$ does not maximize the visibility because heralding increases the effective mean photon number of the signal mode, and thus a lower value of $\mu$ is required to reach maximum visibility compared to two-fold HOM interference.
This effect shifts the diagonal threshold to the left in Fig. \ref{fig:vis_heatmaps}b.

\section{Procedure for fitting HOM interference and teleportation fidelity datasets}
\label{appendix:fitting}

We fit three data sets, two- and three-fold HOM interference visibilities as well as X-basis teleportation fidelity, using a piecewise model function based on our theory. 
Our code performs a nonlinear regression utilizing Mathematica's \texttt{NonlinearModelFit} function with Differential Evolution as the fitting method. 
This global optimization approach is well-suited for fitting nonlinear models. 
As discussed in Sec. \ref{sec:fit}, we utilize six physical parameters: $\eta_s$, $\eta_{i2}$, $\eta_{i3}$, $\mu$, $\zeta_2$, and $\zeta_3$, as fitting parameters, subject to relevant physical constraints. 
Different mode mismatch and signal mode efficiency parameters, $\zeta_2$ and $\zeta_3$, as well as $\eta_{i2}$ and $\eta_{i3}$, respectively, are ascribed to the two- and three-fold detection experiments. 
For the three-fold HOM and teleportation data, a shared $\zeta_3$ is used, with independently determined parameters $\eta_{i3}=1.2\times10^{-2}$, $\eta_{s}=4.5\times10^{-3}$, and $\mu=8.0\times10^{-3}$ from Ref.~\cite{valivarthi2020} remaining constant. 
The two-fold HOM data is fitted independently, retaining $\mu=8.0\times10^{-3}$ as constant.

The fitting protocol is outlined as follows:

\begin{enumerate}
    \item Use the \texttt{Map} and \texttt{Max} functions to ensure uncertainties in the data are bounded to be no less than the square root of the respective y-values.
    \item Combine the three-fold HOM interference and X-basis visibility data, distinguishing them with unique markers. This is achieved using the \texttt{Join} and \texttt{Map} functions.
    \item Formulate a modular fitting function, which can differentiate between X-basis visibility and three-fold HOM interference based on their respective markers. For two-fold HOM data, introduce a separate fitting function that considers the unique constraints of the two-fold detection experiment.
    \item Establish the fitting framework, setting the fitting parameters such as $\eta_{i2}$, $\zeta_2$, and $\zeta_3$ accordingly, while holding the known parameters constant.
    \item Engage in a simultaneous fitting procedure using \texttt{NonlinearModelFit}. This process will take into account the defined model function constraints, weigh the data points based on their squared uncertainties, and adopt the "DifferentialEvolution" fitting technique. The physical constraints on the fitting parameters will ensure that path efficiencies and indistinguishabilities are positive and no larger than unity. 
\end{enumerate}

The outcomes of the fits yield $\zeta_2 = 0.80 \pm 0.04$, $\eta_{i2}=(6.9\pm1.2)\times10^{-2}$, and $\zeta_3 = 0.90 \pm 0.02$ as optimal parameter estimations.

\section{Calculus of HOM interference visibility expressions}
\label{appendix:HOM_calculus}

We differentiate the HOM visibility expressions of Eqs. \ref{eq:2fold_HOMvis} and \ref{eq:3fold_HOMvis} to determine the optimal choice of $|\alpha|^2$.
The expression for the two-fold case (Eq. $\ref{eq:2fold_HOMvis}$), when differentiated with respect to $|\alpha|^2$ and evaluated for the relevant experimental and extracted parameters $\zeta_2= 0.80$, $\eta_{i}=6.9\times10^{-2}$ and $\mu=8.0 \times 10^{-3}$, yields
\begin{align}
    V_{\text{2-HOM}}'(|\alpha|^2) &= \frac{-1.98781 e^{|\alpha|^2/2}-55552.9e^{0.500019|\alpha|^2}+55554.9e^{0.500019|\alpha|^2}}{0.987811 - 1.98781 e^{|\alpha|^2/2} + e^{|\alpha|^2}}.
    \label{eq:2fold_HOMvis_deriv_parameters}
\end{align}

The three-fold case (Eq. $\ref{eq:3fold_HOMvis}$), given a similar treatment with $\zeta_3 = 0.90$, $\eta_i=1.2\times10^{-2}$, $\eta_{s}=4.5\times10^{-3}$ and $\mu=8.0\times10^{-3}$, yields
\begin{align}
   V_{\text{3-HOM}}'(|\alpha|^2) &= 
   \frac{1}{ \left(0.987811-1.98781e^{|\alpha|^2/2}+e^{|\alpha|^2} \right)^2} \Bigl[-0.98179e^{|\alpha|^2/2}-27439.2e^{0.500024|\alpha|^2}+27440.2e^{0.500024|\alpha|^2} \nonumber \\ &+ 2.22045 \times 10^{-16} e^{|\alpha|^2} + 2.63814e^{1.00002|\alpha|^2} \nonumber - 2.65025e^{1.00002|\alpha|^2} + 0.993905e^{3|\alpha|^2/2} \nonumber \\ &+ 27775.1e^{1.50002|\alpha|^2} - 27776.1e^{1.50002|\alpha|^2} \Bigr]
   \label{eq:3fold_HOMvis_deriv}
\end{align}
Setting Eqs. \ref{eq:2fold_HOMvis_deriv_parameters} and \ref{eq:3fold_HOMvis_deriv} equal to zero and evaluating $|\alpha|^2$ results in $7.8 \times 10^{-4}$ and $2.2 \times 10^{-3}$, respectively, which is consistent with the curves shown in Fig. \ref{fig:threefolddataplots}.

\section{X-basis teleportation curves for varying transmission efficiencies and mean photon numbers}
\label{appendix:teleportationcurves}

\begin{figure}[h!]
    \centering
    \includegraphics[width=0.5\textwidth]{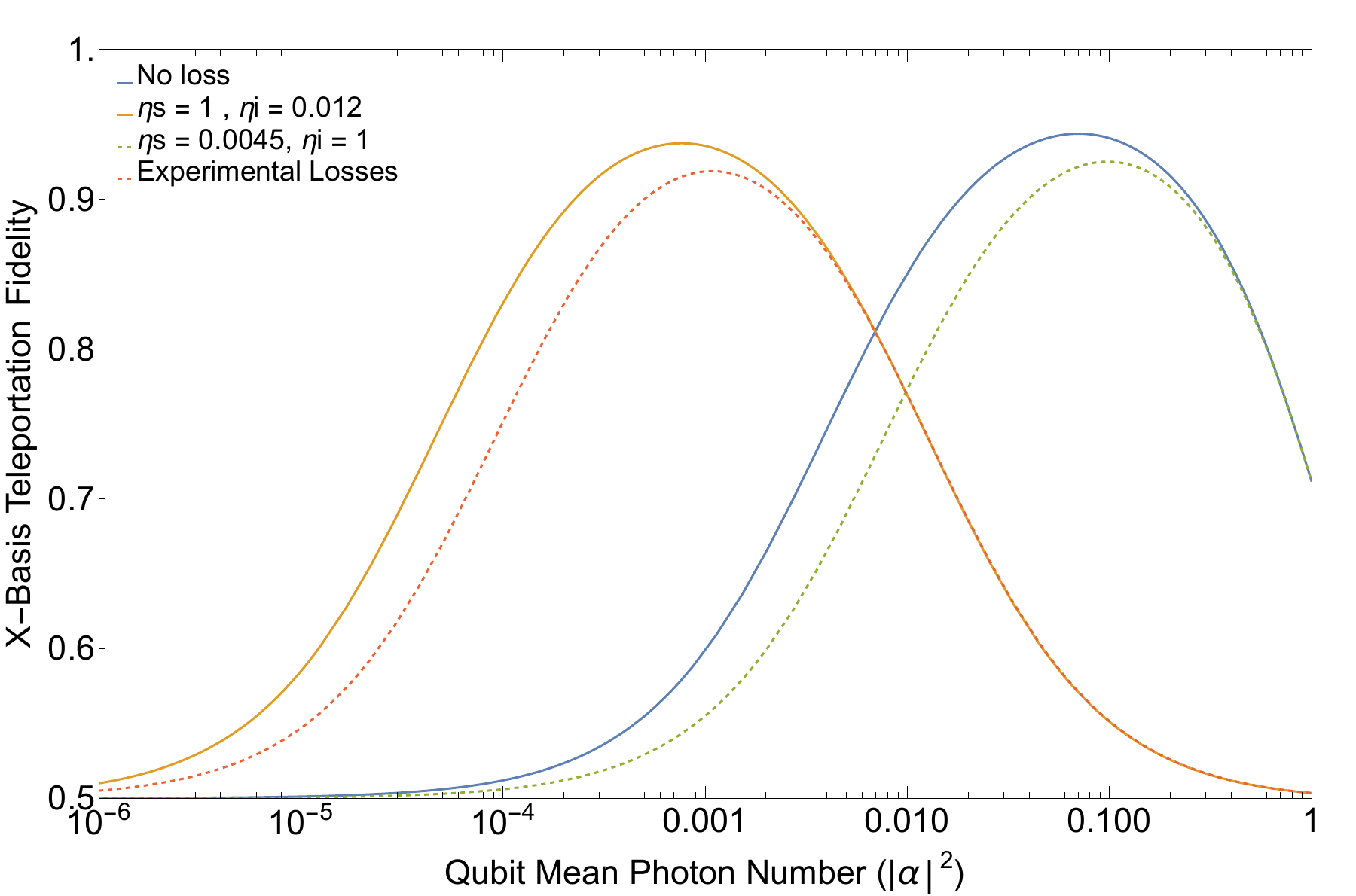}
    \caption{Model of teleportation fidelity of X-basis states for varied $|\alpha|^2$ under conditions of varied signal and idler transmission efficiencies in blue, red, green, and orange, respectively, as described in Sec. \ref{subsection:efficiencies} the main text, assuming complete indistinguishability $\zeta=1$.}
    \label{fig:varyeta_teleportation}
\end{figure}

\begin{figure}[H]
    \centering
    \includegraphics[width=\linewidth]{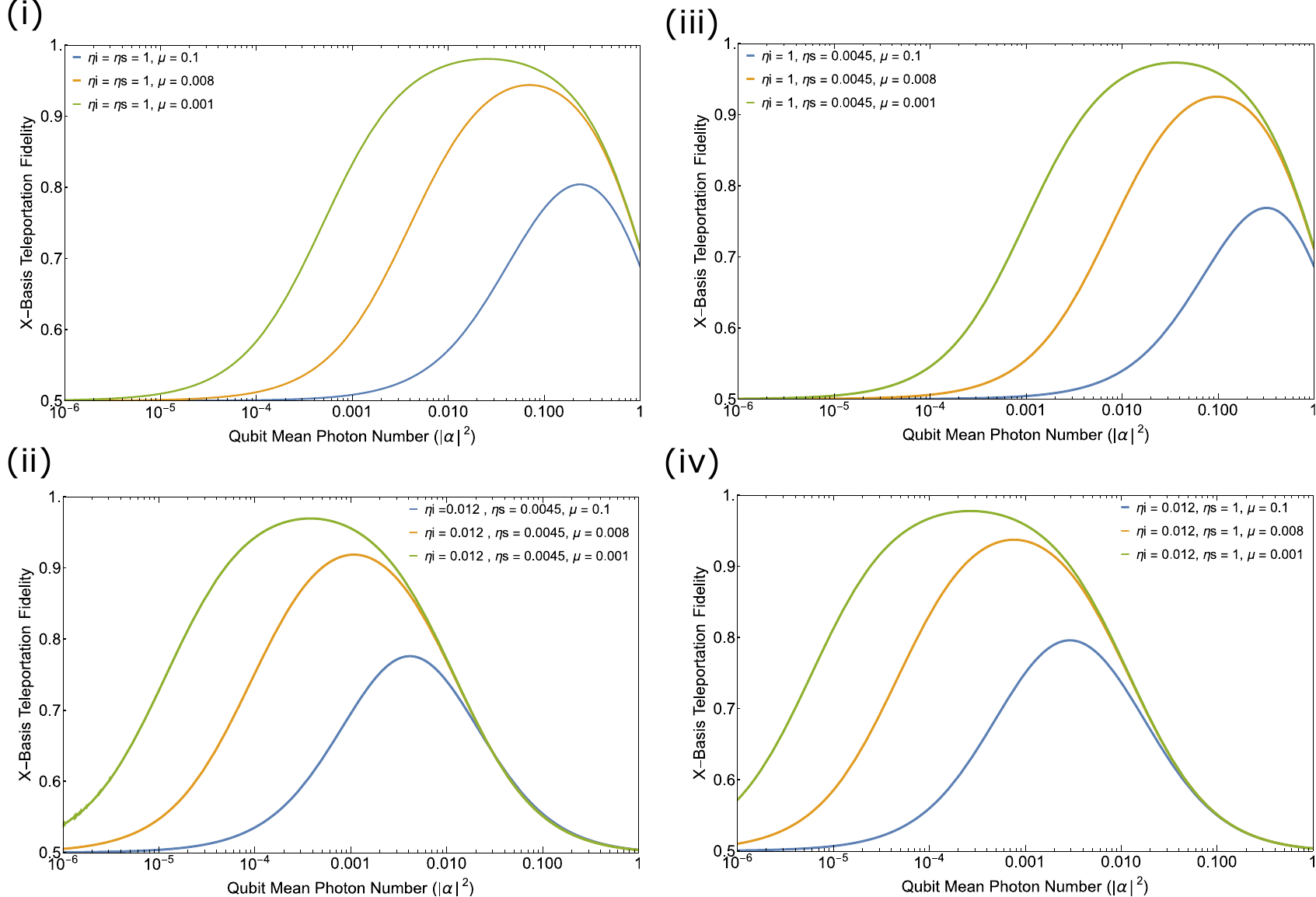}
    \caption{Model of X-basis quantum teleportation fidelity for varied $|\alpha|^2$ and $\mu<10^{-2}$, under varied signal and idler transmission efficiencies cases (i)-(iv), as discussed in Sec. \ref{subsection:mu}, assuming complete indistinguishability $\zeta=1$.}
    \label{fig:varymu_teleportation}
\end{figure}

\twocolumngrid
\bibliography{main}
\end{document}